\documentclass[reprint,nofootinbib,aps,pre]{revtex4-1}
\usepackage[T1]{fontenc}
\usepackage[latin9]{inputenc}
\usepackage{array}
\usepackage{units}
\usepackage{mathtools}
\usepackage{bm}
\usepackage{setspace}
\usepackage{esint}
\usepackage{cancel}
\PassOptionsToPackage{normalem}{ulem}
\usepackage{ulem}
\usepackage[english]{babel}
\usepackage{hyperref}
\usepackage{cleveref}
\usepackage{amssymb}
\usepackage{stmaryrd}
\usepackage{abraces}
\hypersetup{
  colorlinks=true,
  linkcolor=blue,
  filecolor=magenta,
  urlcolor=cyan,
}
\makeatletter

\usepackage{import}
\usepackage{empheq}

\newcommand{\executeiffilenewer}[3]{%
\ifnum\pdfstrcmp{\pdffilemoddate{#1}}%
{\pdffilemoddate{#2}}>0%
{\immediate\write18{#3}}\fi%
}

\newcommand{\executeiffilenewerext}[3]{%
  \ifnum\pdfstrcmp%
    {\pdffilemoddate{#1}}%
    {\pdffilemoddate{#2}}%
    >0%
    {\immediate\write18{#3}}%
  \fi%
}

\newcommand{\eq}{{\rm eq}}
\newcommand{\neqq}{{\rm neq}}
\newcommand\spaciouslvert{\empheqlvert\hspace*{2mm}}
\newcommand{\fZeroEq}{\tilde f_{i}^{{\rm eq},t+1}\left(\bm{x}_{W}\right)}
\newcommand{\fZeroNeq}{\tilde f_{i}^{{\rm neq},t+1}\left(\bm{x}_{W}\right)}
\newcommand{\fOne}{\tilde{f}_{i}^{*}\left(\bm{x}_{W}\right)}
\newcommand{\fOneEq}{\tilde{f}_{i}^{\rm{\eq}}\left(\bm{x}_{W}\right)}
\newcommand{\fOneNeq}{\tilde{f}_{i}^{\rm{\neqq}*}\left(\bm{x}_{W}\right)}
\newcommand{\fZero}{\tilde{f}_{i}^{t+1}\left(\bm{x}_{W}\right)}
\newcommand{\fQplusOne}{f_i^{*}(\bm x_{F})}
\newcommand{\fOneMinusQ}{f_{\bar i}^{*}(\bm x_F)}
\newcommand{\fMinusQ}{f_{\bar i}^{*}(\bm x_{FF})}
\newcommand{\fQ}{f_{i}^{t+1}(\bm x_{F})}
\newcommand{\fMinusQNeq}{f_{\bar i}^{\rm{\neqq}}(\bm x_F)}

\newcommand{\var}{\xi}
\newcommand{\twoldots}{..}

\newcolumntype{H}{>{\setbox0=\hbox\bgroup}c<{\egroup}@{}}
\usepackage[export]{adjustbox}
\usepackage{braket}

\newcommand{\ux}{\bm{x}}

\newcommand{\uxi}{\bm{\xi}}

\newcommand{\uN}{\bm{\nabla}}

\newcommand{\p}{\partial}

\newcommand{\norm}[1]{\left\Vert #1 \right\Vert}
\@ifundefined{showcaptionsetup}{}{%
 \PassOptionsToPackage{caption=false}{subfig}}
\usepackage{subfig}
\makeatother

\begin{document}

\renewcommand{\mathbf}{\bm}
\let\realket\ket
\let\realbra\bra

\global\long\def\bra#1{\realbra{#1}}%

\global\long\def\ket#1{\realket{#1}}%

\global\long\def\eqdef{\overset{{\scriptscriptstyle {\rm def}}}{=}}%
\global\long\def\feqdef{\overset{{\scriptscriptstyle \phantom{{\rm def}}}}{=}}%

\title{Enhanced single-node boundary condition for the Lattice Boltzmann Method}

\author{Francesco Marson}
\email{francesco.marson@unige.ch}\email{marson.francesco@gmail.com}
\author{Yann Thorimbert}
\author{Jonas L\"{a}tt}
\author{Bastien Chopard}
\affiliation{Department of Computer Science, University of Geneva, 1204 Geneva, Switzerland}

\begin{abstract}
 We propose a new way to implement Dirichlet boundary conditions for complex shapes using data from a single node  only, in the context of the lattice Boltzmann method. The resulting novel method exhibits second-order convergence  for the velocity field and shows similar or better accuracy than the well established Bouzidi \emph{et al.}~\citep{bouzidi2001momentum}  boundary condition for curved walls, despite its local nature. The method also proves to be  suitable to simulate moving rigid objects or immersed surfaces either with or without prescribed motion. The core  idea of the new approach is to generalize the description of boundary conditions that combine bounce-back rule with  interpolations and to enhance them by limiting the information involved in the interpolation to a close proximity of the boundary.
\end{abstract}
\maketitle

\section{Introduction}
In recent years, the rising interest in complex flows in numerous applications such as particulate suspensions~\citep{thorimbert_lattice_2018}, porous media~\citep{huber_channelization_2013}, blood flow~\citep{kotsalos_bridging_2019} and multiphase flow~\citep{leclaire_generalized_2017} gave a new impulse to research on local boundary conditions for the Lattice Boltzmann Method (LBM). Local boundary methods for curved geometries can deliver a precise flow description, needing to access the flow variables only on a \emph{single node} located next to the surface. Thanks to these characteristics, it is possible to improve the geometry description, yet maintain an efficient memory access pattern and limit the communications between threads in parallel simulations.

Since the standard lattice Boltzmann method is inherently bounded to its regular and structured lattice, boundary conditions that aim to recover realistic shapes are often of \emph{off-lattice} nature. This implies that some amount of information located outside the concerned mesh node needs to be integrated into the mathematical model, generally using interpolations. If this is not done, the accuracy of the boundary condition degrades to first-order in space, due to the inability to follow the curved shape of the wall.  A first-order representation deteriorates the overall accuracy of the simulation and may require an increase of mesh resolution.  For example, the common \emph{half way bounce-back} rule~\citep{laddNumericalSimulationsParticulate1994}, is an up to third-order accurate scheme~\citep{ginzbourg_local_1996,ginzburg2003multireflection,krugeretal.2016thelattice}, but the solution degenerates to a first-order ``stair-cased'' representation when applied to curved boundaries. In most cases, the use of interpolations or extrapolations causes the loss of locality of the method. Roughly speaking, if we call \emph{boundary nodes} the nodes located next to the surface, the boundary condition will need to access the second layer of nodes, that here we call \emph{secondary} nodes. Nevertheless, in the last three decades some local curved boundary conditions have been proposed~\citep{ginzbourg_local_1996,junk2005onepoint, chun2007interpolated,zhao2017singlenode,tao2018onepoint,liu2019lattice,silva_reviving_2020}.

To cope with the large number of different approaches, it is useful to split boundary techniques into two groups. The first is based on the computation of unknown populations through a unique operation applied to the current node. The second sequentially resolves the unknowns through independent operations in each lattice direction. These last, are often referred to as \emph{link-wise} because they operate on ``links'' that connect the \emph{boundary nodes} with the wall along the discrete lattice directions. For this reason, they do not require any information from the other populations to reconstruct the pre-collision value of the population on a given link. On the contrary, the former are named \emph{node-based}.

A subset of the \emph{link-wise} group consists of techniques inspired by the \emph{half-way bounce-back} (HWBB) rule ~\citep{laddNumericalSimulationsParticulate1994} and commonly referred to as \emph{interpolated bounceback} in recent literature \citep{chun2007interpolated,peng2019acomparative,zhangvelocity,derosis2014acomparison}. The most common HWBB extension to treat curved boundary conditions is the Bouzidi, Firdaouss and Lallemand method (BFL)~\citep{bouzidi2001momentum}. In the (linear) BFL, the populations in an adjacent layer of nodes (\emph{secondary nodes}) are additionally used to carry out interpolations dependent on the wall position. The wall position impacts not only the interpolation coefficients but also the choice of nodes and populations involved in the interpolation scheme. In the present article, we call \emph{fragmented} methods with this propriety. On the contrary, we use the term ``unified'' to refer to algorithms in which the wall position only determines the coefficients of the interpolation, but does not modify the expression interpolation scheme. Two years after the proposition of the BFL, Yu~\citep{yuaunified} proposed a scheme that can be considered as a \emph{unified} version of the linear BFL, given that it uses the same populations as the BFL to perform the interpolation.  The BFL and Yu methods share the drawback of not being local, needing a second layer of nodes to operate.

In the last two decades, few attempts to create local interpolated bounce-back schemes have been proposed~\citep{chun2007interpolated,zhao2017singlenode, tao2018onepoint,liu2019lattice}.
Among those, the Zhao and Yong (ZY)~\citep{zhao2017singlenode} ingeniously chose the LBM populations used to interpolate the unknown mixing pre-collision and post-collision values to build a single-node boundary condition, without the need to introduce further elements to the model. This boundary condition, further developed in references~\citep{zhao_boundary_2019,wang_discrete_2020,meng_simulating_2020}, is second-order accurate in space under diffusive scaling hypothesis and first-order accurate in time~\citep{zhao_boundary_2019}. The ZY method has been tested by Peng \emph{et al.} in~\citep{peng2019acomparative,peng2019acomparative2}. Just like the original paper, they report second-order convergence and accuracy similar to the BFL. The remaining local \emph{interpolated bounce-back} methods~\citep{chun2007interpolated, tao2018onepoint,liu2019lattice} follow a different approach introduced by Chun and Ladd~\citep{chun2007interpolated}: first they reconstruct the boundary populations using the wall velocity and a constant approximation for the density, and then they use an approximated version of the non-equilibrium bounce-back of Zou and He (see~\citep{zou1997onpressure}). However, the method of Chun and Ladd (CL) is not strictly speaking local, as it can require information from other nodes in some situations. In the wake of Chun et Ladd, in recent years two new local boundary conditions have been proposed: the Tao \emph{et al.}~\citep{tao2018onepoint} and the Liu \emph{et al.}~\citep{liu2019lattice}.

Among the first local \emph{link-wise} techniques, the Filippova and Hanel (FH)~\citep{filippova1998gridrefinement,filippova_lattice-boltzmann_1997} can be singled out. The FH belongs to the family of \emph{ghost methods}~\citep{krugeretal.2016thelattice}, also known as \emph{extrapolation methods}~\citep{succi_lattice_2018} or \emph{fictitious equilibrium methods}~\citep{guo_lattice_2013}, that uses additional fictitious nodes on the solid side of the boundary together with extrapolations to reconstruct the unknown populations. In FH the ghost node used for the interpolation is built guessing a velocity beyond the wall with an extrapolation of the boundary velocity. Unfortunately, the FH method has known stability issues, solved by the Mei, Li and Shyy, but sacrificing the locality of the method~\citep{mei1999anaccurate,krugeretal.2016thelattice}.

Junk and Yang (JY)~\citep{junk2005onepoint} proposed a \emph{single node} boundary condition based on a correction of the half-way bounce-back scheme. To perform the correction, it is necessary to solve a linear system on each node to ensure the compliance of the numerical result with the expected solution at Navier-Stokes level. For this reason, it cannot be considered \emph{link-wise}. Besides, the solution of the linear system for each node adds a layer of complexity to the implementation. This approach leads to an almost local mass conservative boundary condition that shows second-order convergence for the velocity. However, the original paper~\citep{junk2005onepoint} reports that the JY method is slightly less accurate than BFL for the velocity and pressure fields in the case of the flow inside a cylinder. The method has been extensively tested by Yang~\citep{yang_analysis_2007} that concluded that the method is almost as stable as the BFL method, and it has comparable or better accuracy of the BFL and FH methods. Nevertheless, the JY method has been tested by Nash \emph{et al.} who reported in~\citep{nash_choice_2014} poor stability properties.

It is worth mentioning two other local non-\emph{link-wise} methodologies. The partially saturated bounce-back (PSBB)~\citep{noble_lattice-boltzmann_1998,noble_lattice-boltzmann_2011} is a local method based on full-way bounce back. The full-way bounce-back rule is based on a modified LBM collision step, while the half-way bounce-back scheme modifies the streaming step. In the PSBB, the underlying idea is to use the knowledge of the fraction of fluid in the boundaries cells to operate a mixed fluid-solid collision~\citep{noble_lattice-boltzmann_2011,krugeretal.2016thelattice}. The PSBB is exactly mass-conservative and allows for a smooth transition between solid and fluid nodes in the case of moving objects. Moreover, it does not require the exact knowledge of the shape of the surface: this is particularly suitable for porous media applications. However, this can turn into a disadvantage when it is necessary to guarantee an exact \emph{no-slip condition} at the surface, because the fluid fraction is not a sufficient piece of information for the method to ``know'' the boundary position and orientation.
 It also requires an additional computation step if the fluid fraction of boundary nodes needs to be recovered from the geometrical shape of the wall. Furthermore, Chen \emph{et al.} comparing different boundary conditions in~\citep{chen2014acomparative} reported a low accuracy in the computation of the cylinder drag when using the PSBB. From the algorithmic point of view, the PSBB cannot be used to represent thin shells because this method constrains the user to allocate solid nodes in the simulation.

 We finally mention the Local Second-Order Boundary (LSOB) method of Ginzburg~\citep{ginzbourg_local_1996}. The technique is based on a precise computation of the boundary nodes according to the Chapman-Enskog expansion to relate macroscopic fields such as density and velocity with the mesoscopic populations of the LBM. The LSOB is a high-fidelity third-order accurate local method, but it is limited to laminar flows~\citep{silva_reviving_2020} and its implementation is lattice and problem-dependent.

 Despite the existence of the single-node boundary conditions that we have reviewed, further research is needed to make local boundary conditions appealing in terms of accuracy and simplicity of the implementation. 
 Furthermore, the relation between the existing local boundary method should be clarified. To this end, in the present article, we develop a framework to generalize interpolated bounce-back schemes, including the CL, Tao, Liu, ZY local methods, and the well-established BFL and Yu methods. Within this framework, we also develop a family of novel boundary conditions to improve the compactness of the interpolation range and the accuracy of the non-equilibrium approximation adopted in the CL, Tao, and Liu methods.

 This article is structured in the following way. After briefly presenting the LBM in \autoref{sec:lbm} and the interpolated bounce-back methods in \autoref{sec:ibb}, a general description of the local ELIBB is presented in \autoref{sec:ELIBB}. In \autoref{sec:ELIBBvariants} some specific variants of the general ELIBB scheme are proposed. Finally, the implementation in the open-source software PALABOS~\citep{latt_palabos_2020} of the ELIBB is tested for three configurations whose analytical solution is known. Namely, the impulsively-started unsteady Couette flow in \autoref{sec:couette}, the steady cylindrical Couette flow in \autoref{sec:couette-cy} and the Jeffery's orbit in \autoref{sec:jeffery}.

\section{Numerical Methodology}
\subsection{The lattice Boltzmann Method}
\label{sec:lbm}
The Boltzmann equation (BE)
\begin{equation}
  \p_{t}f+(\uxi\cdot\uN)f={\cal Q}_{\mathrm{BE}}(f,f)\label{eq:BE}
\end{equation}
describes the space and time evolution of the probability distribution function $f(\ux,\uxi,t)$ of finding a particle
with velocity $\uxi$ at position $\ux$ and time $t$.
The latter is subject to advection in the velocity space ${\bm \xi}$, as well as collision as illustrated by the
Boltzmann's collision integral ${\cal Q}_{BE}(f,f) $~\citep{cercignani_boltzmann_1988,cercignani_mathematical_1994,
cercignani_transport_2007,kremer2010anintroduction}.
The complexity of the Boltzmann's collision integral is the major obstacle to the solution and analysis of the
equation. This is why ${\cal Q}_{BE}(f,f)$ is commonly approximated, with \emph{relaxation towards equilibrium} models.
One of the oldest and most successful relaxation models is the BGK operator that was formulated independently by
Bhatnagar, Gross, Krook~\citep{bhatnagar_model_1954} and by Welander~\citep{welander_temperature_1954}.

In the LB context, the BE (\ref{eq:BE}) is first decomposed in a \textit{finite} set of equations resulting from the velocity space discretization~\citep{SHAN_PRL_80_1998,shan_kinetic_2006}. 
Those equations are known as the Discrete Velocity Boltzmann Equation (DVBE) and read as
\begin{equation}
  \p_{t}f_i+(\uxi\cdot\uN)f_i={\cal Q}_{\mathrm{DVBE}}(f_i,f_i)\label{eq:DVBE}
\end{equation}
Similarly to the BE, the DVBE expresses the time evolution of \textit{discrete} probability distribution functions
$f_{i}(\mathbf{x},\mathbf{\xi}_{i},t)$. But contrarily to the BE, the latter ``populations'' now propagate at
\textit{constant} velocity $\mathbf{\xi}_{i}$ with $i\in \{ 0,1,\ldots,Q\}$.
Applying the method of the characteristics and the trapezoidal integration rule --to the LHS and RHS terms of the
DVBE~(\ref{eq:DVBE}) respectively-- one ends up with the Lattice Boltzmann Equation (LBE) that is the cornerstone of
LBMs~\citep{DELLAR_CMA_65_2013,krugeretal.2016thelattice}
\begin{equation}
  f_{i}(\bm{x}+\bm{c}_{i},\,t+1)=f_{i}\left(\bm{x},t\right)\underset{\mathclap{{\cal Q}_{\text{LBM}}}}{\underbrace{-\Omega f_{i}^{\neqq}}}\label{eq:LBE}
\end{equation}
where lattice units are implied. ${\cal Q}_{\text{LBM}}$ consists in a heuristic approximation of
${\cal Q}_{\text{BE}}$, and $\Omega$ is the relaxation parameter.
$f_{i}^{\eq}$ and $f_{i}^{\neqq}=f_{i}-f_{i}^{\eq}$ are the equilibrium and non-equilibrium populations respectively.
In practice, the LBE~(\ref{eq:LBE}) is solved through two successive steps.
The local \emph{collision} step (RHS of
equation~\eqref{eq:LBE}) and the non-local Lagrangian \emph{streaming} step (LHS of equation~\eqref{eq:LBE}). Through
the normalization of discrete velocities ($\bm{c}_i = \bm{\xi}_i c_s$ with $c_s$ the lattice constant), the streaming
satisfies the on-grid condition that leads to an \textit{exact} advection of populations from one grid node to
another one~\cite{krugeretal.2016thelattice}.
To compute the collision term, one needs to derive equilibrium and non-equilibrium populations. 
$f_i^{\eq}$ is the discrete counterpart of the Maxwell-Boltzmann distribution~\citep{kremer2010anintroduction}
\begin{equation}\label{eq:Maxwellian}
  f^{\eq}=\frac{\rho(\boldsymbol{x}, t)}{(2 \pi \theta(\boldsymbol{x}, t))^{D
  / 2}} \exp \left(-\frac{(\boldsymbol{u}(\boldsymbol{x}, t)
  -\boldsymbol{\xi})^{2}}{2 \theta(\boldsymbol{x}, t)}\right)\, ,
\end{equation}
where $D$ is the dimensionality of the problem. $\rho$, $\bm{u}$ and $\theta$ are the macroscopic density, velocity and reduced temperature respectively.
One way to derive $f_i^{\eq}$ from the Maxwellian~(\ref{eq:Maxwellian}) is to rely on the Gauss-Hermite quadrature~\cite{SHAN_PRL_80_1998,shan_kinetic_2006}. In the context of isothermal LBMs, the reduced temperature $\theta=T_0$ is constant and directly related to the lattice constant $T_0=c_s^2$, where $c_s^2=1/3$ for standard velocity discretizations (D2Q9, D3Q27, etc). 

Regarding collision models, one of the most popular is the BGK approximation that leads to~\citep{qian_lattice_1992}
\begin{equation}
  \Omega_{\text{BGK}}=\frac{1}{\bar{\tau}+1/2},\qquad f_{i}^{\neqq}=\left(f_{i}-f_{i,2}^{\eq}\right)
  \label{eq:LB-BGK}
\end{equation}
where $\bar{\tau}$ is the relaxation time, even though, in the LB community, it is also common to refer to $\tau=\bar{\tau}+1/2$ as relaxation time.
\begin{equation}
f_{i,2}^{\eq}=w_{i} \rho\bigg[1+ \dfrac{c_{i\alpha_1}u_{\alpha_1}}{c_{\mathrm{s}}^{2}}+ \dfrac{u_{\alpha_{1}}u_{\alpha_{2}}(c_{i\alpha_{1}}c_{i\alpha_{2}}- c_{\mathrm{s}}^{2}\delta_{\alpha_{1}\alpha_{2}})}{2c_{\mathrm{s}}^{4}}\bigg]
  \label{eq:Eq2ndOrder}
\end{equation}
is the discrete equilibrium up to the second-order, where $\alpha_i\in\{x_1,\ldots,x_d\}\ \forall\ i$ and Einstein's summation rule is assumed for the sake of compactness. The BGK collision model is easy to understand and to implement, but it has several drawbacks. Concerning the bounce-back method, the coupling with the BGK collision model gives rise to a \emph{second-order} error (related to the exact location of the wall) that is commonly referred to as \emph{viscosity-dependent error}. Among the other limitations, it can lead to numerical instabilities in the case of under resolved meshes (typically at high Reynolds numbers) and non-vanishing Mach numbers~\citep{krugeretal.2016thelattice,succi_lattice_2018,COREIXAS_RSTA_378_2020}.

To overcome these issues, several extended collision models were proposed~\cite{coreixas2019comprehensive} with varying degrees of success~\cite{COREIXAS_RSTA_378_2020}. As a first study, hereafter, we will restrict ourselves to Multi-Relaxation-Time (MRT)~\citep{dhumieres_generalized_1992,LALLEMAND_PRE_61_2000,dhumieres_multiplerelaxationtime_2002}, Two-Relaxation-Time (TRT)~\citep{ginzburg_two-relaxation-time_2008,ginzburg2008studyof,DHUMIERE_CMA_58_2009,GINZBURG_JSP_139_2010}, and regularization~\citep{latt2006lattice,Malaspinas2015,coreixas_recursive_2017,JACOB_JT_19_2018,WISSOCQ_ARXIV_2020_07353} approaches, which all have a different impact on the boundary condition performances~\cite{SILVA_PRE_96_2017}.

\paragraph{MRT models:}
The core idea of the MRT is to carry out the relaxation collision process in the moment space. In other terms, from $Q$ populations, $Q$ moments are computed.
The latter is (independently) relaxed towards their equilibrium values using different relaxation parameters.
Referring to equation (\ref{eq:LBE}), we can mathematically describe the MRT collision model as follows
\begin{equation}
  \bm{\Omega}_{\text{MRT}}=\bm{M}^{-1}\bm{S}\bm{M},\qquad f_{i}^{\neqq}=\left(f_{i}-f_{i,2}^{\eq}\right)
\end{equation}
where, in the original formulations, $\bm S$ is a diagonal relaxation matrix, while $\bm M$ and $\bm M^{-1}$ are two orthogonal matrices allowing to move from the Q-dimensional velocity space to the Q-dimensional orthogonal moment space and \emph{vice versa}.  In this context, the LBE (\ref{eq:LBE}) can be rewritten for the MRT case using the \emph{bra-ket} notation
\begin{multline}
  \ket{f_{i}(\bm{x}+\bm{c}_{i},\,t+1)}=\\
  \ket{f_{i}^{\eq}}+\left(\bm{I}-\bm{M}^{-1}\bm{SM}\right)\ket{f_{i}^{\neqq}}
  \label{eq:lb-mrt}
\end{multline}
where the symbol $\ket{\ldots}$ stands for the vectorial notation populations. This method depends on many free parameters, though, which need to be handpicked through an adequate procedure.

\paragraph{TRT models:}
To reduce the number of free parameters, Ginzburg \emph{et al.} \citep{ginzburg_two-relaxation-time_2008,ginzburg2008studyof,DHUMIERE_CMA_58_2009,GINZBURG_JSP_139_2010} proposed a two-relaxation formulation of the MRT collision model~(\ref{eq:lb-mrt}). These TRT-LBMs independently relax even (symmetric) and odd (anti-symmetric) moments, each of which are assigned an individual collision frequency: $\omega^{+}$ and $\omega^{-}$ respectively.
Interestingly, the TRT   behavior can be described by the \emph{magic parameter} $\Lambda$ that relates the two collision frequencies in the following way\begin{equation}
  \Lambda=\left(\frac{1}{\omega^{+}}-\frac{1}{2}\right)\left(\frac{1}{\omega^{-}}-\frac{1}{2}\right).
  \label{eq:magic-parameter}
\end{equation}
TRT-based collision models are particularly interesting for the bounce-back boundary
condition.
It has been proven that for $\Lambda=\nicefrac{3}{16}$ the bounce-back condition is viscosity-independent meaning, i.e., the wall location does not depend on the viscosity anymore. The latter feature is particularly critical for low-viscosity flows, and it further leads to third-order accuracy for the half-way bounce back~\citep{krugeretal.2016thelattice}.

\paragraph{Standard and recursive regularized models:}
The instabilities of the BGK model can be addressed in a way that is an alternative to the MRT and TRT approach (even if can be reformulated in the same formalism for particular moment spaces and relaxation frequencies~\cite{coreixas2019comprehensive,COREIXAS_RSTA_378_2020}).
The basic idea of the regularized (RBGK) and the recursive regularized (RRBGK) models is to filter out non-hydrodynamic modes in the BGK solution arising from the numerical discretization of the velocity space. The latter spurious contributions are hidden in non-equilibrium populations and can be deleted by ``manually'' imposing a particular form --compliant with the macroscopic  behavior of interest-- of $f_i^{\neqq}$. More precisely, starting from a Knudsen number expansion of populations (as done, e.g., for any Chapman-Enskog expansion), one obtains
\begin{equation}\label{eq:CE}
    f_i = \sum_{l=0}^{\infty} f_i^{(l)} = f_i^{(0)} + f_i^{(1)} + f_i^{(2)} + \ldots
\end{equation}
with $f_{i}^{(l)}\sim \mathcal{O}(\mathrm{Kn}^l f^{(0)})$ the $l$-order contribution (with respect to the Knudsen
number Kn) to populations $f_i$, and $f_{i}^{(0)}=f_i^{\eq}$~\cite{chapman_mathematical_1953}.
Each regularization step aims at manually discarding terms that are not compliant with the Navier-Stokes-level of physics, i.e.,
\begin{equation}
  f_i^{\neqq} = f_i - f_{i}^{(0)} \approx f_{i}^{(1)}
\end{equation}
where $f_{i}^{(1)}$ can have several forms depending on the assumption of the considered regularized approach~\cite{latt2006lattice,Malaspinas2015,coreixas_recursive_2017,JACOB_JT_19_2018,WISSOCQ_ARXIV_2020_07353}.
An in-depth discussion regarding that point can be found in \cref{sec:reg} and \cref{sec:hydro_limit}.

\subsection{Interpolated bounce-back methods}
\label{sec:ibb}
\begin{figure}[bp!]
  \includegraphics[scale=1.1]{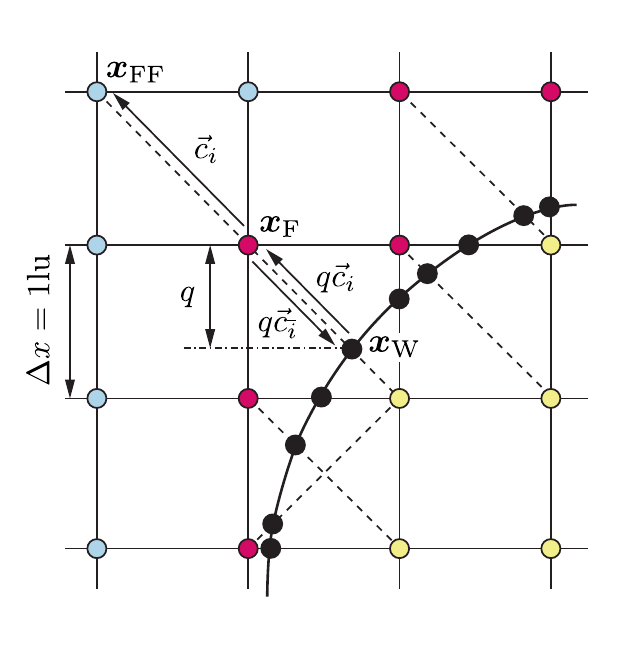}
  \caption{2D representation of the boundary nodes, normalized distance $q$,
  discrete lattice velocities $\vec c_i,\,\vec c_{\bar i}$, links (dashed segments)
  and locations of boundaries at the intersections with links ($\bullet$). $\Delta x$ is the space step.}
  \label{fig:2d-boundary}
\end{figure}
The purpose of boundary algorithms is to reconstruct missing populations  on nodes next to the wall (\emph{boundary nodes}), after the streaming step.  On the \emph{boundary nodes} F, the unknown population are  those associated with discrete velocities  $\bm c_i$  that ``leave'' the wall (see figure~\ref{fig:2d-boundary}).  We call such populations
``\emph{incoming to fluid}'' and denote them with index $i$,
while we call ``\emph{outgoing from  fluid}'' and denote with the index $\bar i$ the others~\citep{succi_lattice_2018}.  The main idea of the Interpolated Bounce-Back methods (IBB) is to perform  a one dimensional polynomial interpolation of the known population close  to the boundaries along the discrete directions of the lattice (links directions) to recover the  unknown \emph{incoming} populations at the \emph{boundary nodes}.  We can formalize this general idea in mathematical terms writing
\begin{subequations}
  \begin{equation}
    f_{i}(\bm{x}_{F},t+1)=\sum_{j}a_{j}f_{i}^{*}(\bm{x}_{j},t)+\sum_{k}a_{k}f_{\bar{i}}^{*}(\bm{x}_{k},t)+K
    \label{eq:general-ibb}
  \end{equation}
where $f^*=f-\Omega f^\neqq$ is the post-collision population, $K$ is a hypothetical correction factor, the symbol $a$ denotes the interpolation coefficients, $\bm x$ is an interpolation point and $t$ is the current iteration.    In practice, for the linear    case and referring to figure~\ref{fig:2d-boundary}, the previous formula    generally reduces to
  \begin{align}
    f_{i}(\bm{x}_{F},t+1)= & a_{1}f_{\bar{i}}^{*}(\bm{x}_{FF},t)+a_{2}f_{\bar{i}}^{*}(\bm{x}_{F},t)\nonumber\\
    & +a_{3}f_{i}^{*}(\bm{x}_{F},t)+K
    \label{eq:general-ibb-linear}
  \end{align}
  \label{eq:ibb}
\end{subequations}
equations~\eqref{eq:ibb} represent a generic formulation of the interpolated-bounceback approach. To derive a specific method, it is necessary to specify the expressions of the interpolation coefficients and points. To this end, two viable solutions exist. The first one consists in writing closure relations by exploiting the macroscopic \emph{no-slip} condition~\citep{ginzbourg1994boundary,ginzburg2003multireflection,ginzburg2008tworelaxationtime,zhao2017singlenode}. In practice, the no-slip condition is expanded using a formal mathematical expansion and subsequently equations~\eqref{eq:ibb} are injected onto it. The second solution relies on a mesoscopic, geometrical approach and was proposed by Bouzidi~\emph{et al.} among other authors~\citep{bouzidi2001momentum,yuaunified,chun2007interpolated,tao2018onepoint}. The idea in this case is to use the bounce-back rule, understood as a modification to the streaming step, to compute the interpolation coefficients. Roughly speaking, the bounce-back operator modifies the streaming operator from a simple translation in space to a translation-reflection-translation. The populations subjected to the bounce-back, during the translation, are reflected when they encounter the wall. Owing to this bounce-back rule, the interpolation coefficients are those that allow to geometrically compute the unknown either at time step $t+1$ or at its virtual off-lattice post-collision state at time $t$.

\begin{figure}
  \centering
  \includegraphics[width=.99\columnwidth,valign=c]{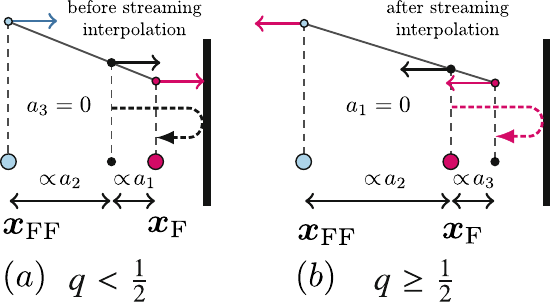}
  \caption{1D link-wise representation of bouncing-back procedure in the BFL method: (left) $q < 1/2$ and (right) $q \geq 1/2$.
  The coefficients $a_1,a_2,a_3$ are defined in equations~\eqref{eq:bfl}. The arrows represent the
  populations appearing in equations~\eqref{eq:bfl}. The dashed curved arrows
  represent the bounce-back rule. Left: the streaming and the bounce-back rule are applied after
  the interpolation of the off-lattice population. Right: the streaming and the bounce-back rule
  are applied before the interpolation, that is carried out at timestep $t+1$.
  \label{fig:bouzidi}}
\end{figure}

To illustrate this concept we consider the BFL algorithm~\citep{bouzidi2001momentum}. For the linear BFL algorithm, equation~\eqref{eq:general-ibb-linear} becomes \begingroup\allowdisplaybreaks
\begin{subequations}
  \label{eq:bfl}
  \begin{align}
    f_{i}^{t+1}(\boldsymbol{x}_{F})= &
    \aoverbrace[L1R]{2q}^{\mathclap{a_{1}}}f_{\bar{i}}^{*}(\boldsymbol{x}_{F})+\aoverbrace[L1R]{
    (1-2q)}^{a_{2}}f_{\bar{i}}^{*}(\boldsymbol{x}_{FF}) & q<0.5\\
    f_{i}^{t+1}(\boldsymbol{x}_{F})= &
    \aunderbrace[l1r]{\frac{1}{2q}}_{a_{1}}f_{\bar{i}}^{*}(\boldsymbol{x}_{F})
    +\aunderbrace[l1r]{\frac{2q-1}{2q}}_{a_{3}}f_{i}^{*}(\boldsymbol{x}_{F}) &
    q\geq0.5
  \end{align}
\end{subequations}
\endgroup
where $q$ denotes the distance of the boundary node F from the wall, normalized by the norm of the discrete velocity $\norm{\bm c_i}=\norm{\bm x_{FF}-\bm x_F}$. As showed by figure~\ref{fig:bouzidi}, in the IBB methods, populations can be thought as lumped mass elements moving according to their discrete lattice velocities $\bm c_i$. During the streaming step, each population undergoes either to the free streaming (along a straight line) or to the bounce-back streaming (in case of a wall encounter). In the latter case, the populations revert their streaming direction before completing their trajectory of the length of $\norm{\bm c_i}$. When $q < 1/2$ (figure~\ref{fig:bouzidi}a), the interpolation is carried out when all populations are at the time-step $t$ in their post-collision state. In this case, the target location of interpolation is the former position of $f_{i}^{t+1}(\boldsymbol{x}_{F})$ \emph{before} the bounce-back streaming step. On the contrary, when $q\geq 1/2$ (figure~\ref{fig:bouzidi}b), the interpolation factors must be computed \emph{after} the streaming procedure. Consequently, either the streaming or the bounce-back rule is applied to each population at time-step $t$ (post-collision state). After that, the interpolation is carried out at time-step $t+1$.

\subsection{Single-node interpolated bounce-back methods and the new ELIBB scheme}
\label{sec:ELIBB}
In order to design local link-wise BCs, one must discard the \emph{non-local} contribution $f^*(\bm x_{FF})$ that appears in equations~\eqref{eq:general-ibb-linear} and~\eqref{eq:bfl}. From the algorithmic point of view, one could simply apply the boundary method after streaming and save the other population at node F before the streaming. Unfortunately, this algorithmic locality leads to some issues. In fact, it does not allow to describe corners or narrow gaps (where a second wall is located between the nodes F and FF without introducing a special treatment of these cases~\citep{chun2007interpolated}). To get rid of the unwanted $f^*(\bm x_{FF})$ population, yet maintaining the \emph{link-wise} nature of the method, two approaches have been proposed. The first one comes from Zhao \emph{et al.}~\citep{zhao2017singlenode,zhao_boundary_2019} and consists on the following first-order in time approximation
\begin{equation}
  f_{\bar i}^*(\bm x_{FF}) = f_{\bar i}^{t+1}(\bm x_F) \approx f_{\bar i}^{t}(\bm x_F)\, .
\end{equation}
The second one, was introduced by Chun and Ladd~\citep{chun2007interpolated}, further developed by Tao, Liu \emph{et al.}~\citep{tao2018onepoint} and also improved by the present article. It consist building virtual (approximated) population located at the wall position $\bm x_W$. In the proposition of Tao \emph{et al.}~\citep{tao2018onepoint}, only the population $\tilde{f}_i^{t+1}(\bm x_W)$ was introduced in equation~\eqref{eq:general-ibb-linear}, where the $\sim$ cap indicates an approximated value. Hereafter, we propose a novel Enhanced Local Interpolated Bounce-Back method (ELIBB) which additionally accounts for the population $\tilde{f}_i^*(\bm x_W)$ hence, extending~\eqref{eq:general-ibb-linear} in the following way
\begin{empheq}[left=\rotatebox{90}{\hspace*{-3mm}$\rm ELIBB$}\empheqlvert]{align}
  f_{i}\left(\bm{x}_{F},t+1\right)= & \cancel{a_{1}f_{\bar{i}}^{*}\left(\bm{x}_{FF}\right)}+a_{2}\underbrace{f_{\bar{i}}^{*}\left(\bm{x}_{F}\right)}_{\text{Ladd's BB}}+\nonumber\\
  & a_{3}f_{i}^{*}\left(\bm{x}_{F}\right)+a_{4}\underbrace{\tilde{f}_{\bar{i}}\left(\bm{x}_{W},t+1\right)}_{\text{by Tao et al.}}+\nonumber\\
  & a_{5}\underbrace{\tilde{f}_{\bar{i}}^{*}\left(\bm{x}_{W}\right)}_{\text{new ELIBB}}.
  \label{eq:ELIBB-general}
\end{empheq}
The previous equation represents the scheme underlying the new ELIBB method. However, it is still necessary to specify the expression of the populations and the interpolation coefficients. It will be done in the following~\cref{sec:general-interpolation}.
Changing the interpolation coefficients in equation~\eqref{eq:ELIBB-general}, different variants will be developed (\cref{sec:ELIBBvariants}). From this perspective, the model proposed by Tao \emph{et al.}~\citep{tao2018onepoint} directly flows from the ELIBB (equation~\eqref{eq:ELIBB-general}).

\subsubsection{Approximation of the wall populations}\label{sec:approx-wall-pop}
In the approach introduced by Chun in~\citep{chun2007interpolated}, the fundamental idea to realize a single-node IBB is to exploit the knowledge of the boundary velocity to rebuild some \emph{virtual populations} at the boundary location. This approach consists of using the boundary velocity and an estimation of the density at the boundary to rebuild the equilibrium population. In~\citep{chun2007interpolated}, it has been proved that for a ``slow'' flow the approximation
\begin{align}
  \fZeroEq & \approx f^{\eq}\left(\rho(\bm{x}_{F},t),\bm{u}(\bm{x}_{W},t+1)\right)
  \label{eq:feq_approx-1}
\end{align}
is second order accurate. Regarding the approximation of the \emph{non-equilibrium} component  at the wall $\fZeroEq$ we can use an \emph{approximated non-equilibrium bounce-back} (\citep{chun2007interpolated,tao2018onepoint,liu2019lattice}). This is a first-order approximation of the \emph{non-equilibrium bounce-back} method of Zou and He~\citep{zou1997onpressure}. We discuss it in \cref{sec:hydro_limit} [see in particular equation~ \eqref{eq:approx_neq_bb}]. This leads to the following second order accurate approximation
\begin{equation}
  \fZeroNeq \approx \fMinusQNeq
  \label{eq:neqbb-s}
\end{equation}
As demonstrated in~\citep{chun2007interpolated} the reason why the approximated non-equilibrium bounce-back leads to a second-order accurate boundary condition resides in the fact that the non-equilibrium component is a second-order correction over the equilibrium. This fact allows for a second-order approximation with an only first-order approximation of the non-equilibrium part.

In the present article, we additionally propose a more general approach to estimate the non-equilibrium component. The idea is to use Malaspinas'~\citep{Malaspinas2015} recursive formulas to recompute the wall non-equilibrium component using the Hermite basis expansion truncated up to the fourth-order (refer to \cref{sec:hydro_limit} for the details of this procedure). This approach allows for higher flexibility in the modeling of the wall non-equilibrium population.

\subsubsection{Generalized computation of interpolated coefficients in IBB methods}
\label{sec:general-interpolation}
To obtain a more uniform picture, we propose to express the populations in all these IBB methods in their \emph{pre-collision} state at time $t+1$ in a similar fashion of the case  $q\geq1/2$ of the BFL method (figure~\ref{fig:bouzidi}).  To be able to extend this description to all methods and for any value of $q$, we introduce signed normalized distance from the wall $s$ at time $t+1$.  At the time $t+1$ some population has been streamed following the free stream rule,  whilst, others near the wall have been streamed using the bounce-back rule described in \cref{sec:ibb} and figure~\ref{fig:bouzidi}b.  In this condition the generalized coordinate $s$ reads
\begin{equation}
  s\left(f_{{\cal I}}(\bm{x},t+1)\right)\eqdef\frac{(\bm{x}-\bm{x}_{W})\cdot\bm{c}_{{\cal I}}}{\norm{\bm{c}_{i}}}\quad\forall\:{\cal I}\in\{i,\bar{i}\},
  \label{eq:s}
\end{equation}
where $\bm x$ is the coordinate of the population $f_{{\cal I}}$ after the streaming/bounce-back step.
The coordinate $s$ turns out to be a simple yet effective tool to describe and compare link-wise boundary conditions. Using equation~\eqref{eq:s}, we can define a set of simple rules to move from the $\bm x$ coordinate metric to the $s$ coordinate metric:
\begingroup \allowdisplaybreaks
\begin{subequations}
  \begin{empheq}[]{flalign}
    &f_i^{t+1}(\bm x_F) \equiv f(s=q) \\
    &f_i^{t,*}(\bm x_F)=f_i^{t+1}(\bm x_{FF})\equiv  f(s=q+1)     \\
    &f_{\bar i}^{t,*}(\bm x_F) \equiv f(s=-q+1) \\
    &f_{\bar i}^{t,*}(\bm x_{FF})=f_{\bar i}^{t+1}(\bm x_{F}) \equiv f(s=-q)  \\
    &f_{\bar i}^{t,*}(\bm x_W)\equiv f(s=1+\bm u_w \cdot \bm c_i / \norm{\bm c_i})   \\
    &f_{\bar i}^{t+1}(\bm x_W)\equiv f(s=0)\, ,
  \end{empheq}
  \label{eq:x_to_s_helper}
\end{subequations}\endgroup
where the index of $\bm x$ indicates either a node (F or FF) or the virtual node W located at the intersection of the link with the boundary surface (as a consequence $f(\bm x_{\rm W})$ are some virtual population which will be useful later) and $\bm u_w$ is the wall velocity (in lattice units).

With the help of the coordinate $s$, it is now possible to generalize the formulas used in the BFL and the other IBB methods with the Sylvester-Lagrange~\citep{de_lagrange_cons_1812,sylvester_xxxix_1883} polynomial interpolation formula:
\begin{equation}
  f\left(s_{ref},\,t+1\right)=\sum_{j=0}^{n}a_{j}(s_{j})f^{\left(s_{j},\,t+1\right)}
  \label{eq:interpolated_bb_equation}
\end{equation}
where $n$ is the interpolation order, $j$ is the index of the interpolation point, and $a_j$ are the interpolation coefficients given by
\begin{equation}
  a_i=\prod_{\stackrel{\!0\leq j\leq n}{j\neq i}}\frac{s_{ref}-s_j}{s_i-s_j}\ .
  \label{eq:poli_coeff}
\end{equation}
In the linear case the interpolation coefficients $a_{j}\in\{a_\alpha(s_\alpha),a_\beta(s_\beta)\}$ can be easily recovered from the values of $s_{j}\in\{s_\alpha,s_\beta\}$ in the following way
\begingroup
\allowdisplaybreaks
\begin{subequations}
  \begin{align}
    a_{\alpha}\left(s_{\alpha}(q)\right) & =1-\frac{s_{ref}-s_\alpha}{s_\beta-s_\alpha}\label{eq:a1}\\
    a_{\beta}\left(s_{\beta}(q)\right) & =\frac{s_{ref}-s_\alpha}{s_\beta-s_\alpha}\label{eq:a2}\\
    s_\alpha<  s_{ref} &< s_\beta
  \end{align}
  \label{eq:interpolation-factors}
\end{subequations}
\endgroup
where  in our case $s_{ref}=q$. equations~\eqref{eq:interpolation-factors} will be used to develop variants of the general ELIBB formula (equation~\eqref{eq:ELIBB-general}).

\subsection{Enhanced Single-Node boundary condition variants}
\label{sec:ELIBBvariants}
The novel generalized coordinate introduced in the previous section (equation~\eqref{eq:s}) is a useful tool to develop variants of the general scheme proposed in equation~\eqref{eq:ELIBB-general}. The first method that we propose is a \emph{unified} method, and it can be written as:
\begin{subequations}
  \begin{equation}
    \text{ELIBB-U | } \fQ=a_{4}\fZero+a_{5}\fOne
    \label{eq:unified_space_coordinates}
  \end{equation}
  where $\tilde{f}$ is the approximation of population $f$ consisting   of a separate evaluation of the equilibrium $\tilde{f}^{{\rm eq}}$   and non-equilibrium $\tilde{f}^{{\rm neq}}$ parts that was discussed in \autoref{sec:approx-wall-pop}.
 As anticipated in the introduction, the adjective \emph{unified} refers to the fact that the interpolation scheme does not depends on the value of $q$. The interpolation factors $a_{4,5}$ in equation~\eqref{eq:unified_space_coordinates} can recovered converting the $x$ coordinates in $s$ coordinates with equations~\eqref{eq:x_to_s_helper} and then using equations~\eqref{eq:a1} and~\eqref{eq:a2}
  \begin{align}
    a_4 &= 1-q\\
    a_5 &= q \,.
  \end{align}
\end{subequations}

For the wall populations we adopted the following approximations
\begin{subequations}
  \begingroup
  \allowdisplaybreaks
  \begin{empheq}{align}
    \fOne & \eqdef\fOneEq+\fOneNeq\label{eq:ELIBB-approximations}\\
    \fZero & \eqdef\fZeroEq+\fZeroNeq\label{eq:f=eq+neq}\\
    \fZeroEq & \approx f^{\eq}\left(\rho_{F}(t),\bm{u}(\bm x_W,t+1)\right)\label{eq:ELIBB-approximation-eq1}\\
    \fOneEq & \approx f^{\eq}\left(\rho_{F}(t),\bm{u}(\bm x_W,t)\right)\label{eq:ELIBB-approximation-eq2}\\
    \tilde{f}_i^{\rm neq}(\bm x_W)
    &\approx\text{ use }\eqref{eq:neqbb-s}
    \text{ or }\eqref{eq:fi_neq},
  \end{empheq}
  \endgroup
\end{subequations}
where in~\eqref{eq:ELIBB-approximation-eq1} and~\eqref{eq:ELIBB-approximation-eq2} we used~\eqref{eq:feq_approx-1}. In this regard, one may note that in general $\bm x_W(t)\neq \bm x_W(t+1)$ and $\bm u(\bm x_W(t))\neq \bm u(\bm x_W(t+1))$, but their actual expressions depend on the time advance scheme of the wall (\emph{e.g.} explicit euler, implicit euler, \emph{etc.}). Nonetheless, if $\Delta \bm u_w \ll \bm u_w$ one can set $\bm u(\bm x_W(t))\approx \bm u(\bm x_W(t+1))$.
This first variant of the ELIBB method is different from the other because it does not use any nodal population in the interpolation. In particular, if the non-equilibrium component at the wall is computed from a \emph{regularization procedure} (\emph{e.g.} equation~\eqref{eq:fi_neq}) instead of using equation~\eqref{eq:neqbb-s}, a good name for this method could be \emph{local regularized link-wise method}. This name emphasizes the fact that this specific variant only relies on the reconstruction of the wall populations just like the regularized boundary condition presented in the articles~\citep{latt_straight_2008,malaspinas2011general}.

The physical locality of (\ref{eq:unified_space_coordinates}) can be improved adopting a \emph{fragmented} interpolation scheme, taking advantage of the knowledge of the population $\fOneMinusQ$,
\begin{subequations}
  \begin{empheq}[left=\rotatebox{90}{\hspace*{-6mm} \rm ELIBB-F}\spaciouslvert]{align}
    \fQ=&\phantom{+}a_{4} \fZero \nonumber\\&+a_{2}\fOneMinusQ & q<0.5\\
    \fQ=&\phantom{+}a_{2}\fOneMinusQ\nonumber\\&+a_{5} \fOne & q\geq0.5
    \label{eq:linear_broken_method}
  \end{empheq}
  where:
  \begin{align}
    a_2 &= (1-2q)/(1-q)     & q<0.5\\
    a_5 &= q/(1-q)          & q<0.5\\
    a_2 &= (2-2q)/q         & q\geq 0.5\\
    a_5 &= (2q-1)/q         & q\geq 0.5
  \end{align}
\end{subequations}

Finally, the combination of all the information of the two previous methods, it is possible to perform a quadratic interpolation. That leads to a quadratic interpolation unified method
\begingroup
\allowdisplaybreaks
\begin{empheq}[left=\rotatebox{90}{\hspace*{-7mm}\rm ELIBB-UQ}\spaciouslvert]{align}
  \fQ=&a_{4}\fZero \nonumber\\
  +&a_{2} \fOneMinusQ \nonumber \\
  +&a_{5}\fOne
  \label{eq:elibbqu}
\end{empheq}
\endgroup
where the interpolation coefficients $a_i$ are computed using equations~\eqref{eq:poli_coeff} (after using~\eqref{eq:x_to_s_helper} to compute the $s$ coordinates).

Adding the knowledge of $\fOne$ we can also formalize a  fragmented quadratic method
\begin{subequations}
  \begingroup
  \allowdisplaybreaks
  \begin{empheq}[left=\rotatebox{90}{\hspace*{-7mm} \rm ELIBB-FQ}\spaciouslvert]{align}
    \fQ=&a_{4}\fZero & q<0.5\nonumber\\
    +&a_{2} \fOneMinusQ \nonumber \\
    +&a_{5}\fOne \\
    \fQ=&a_{2}\fOneMinusQ& q>0.5\nonumber\\
    +&a_{5}\fOne\nonumber\\
    +&a_{3}\fQplusOne
  \end{empheq}
  \endgroup
  \label{eq:quadratic_fragmented_method}
\end{subequations}
where the interpolation coefficients are given again by equation~\eqref{eq:poli_coeff}. The above quadratic interpolation formulas defined by equations~\eqref{eq:quadratic_fragmented_method} and~\eqref{eq:elibbqu} improve the accuracy of the ELIBB-F. Nevertheless, it is necessary to specify a \emph{cutoff} minimal value $q^{\rm min}_{\rm cutoff}$ below which the ELIBB-F is used in place of the ELIBB-FQ. Thereby, we avoid instabilities due to the overlapping of interpolation points when $q$ approaches zero. For the results presented in this article we use $q^{\rm min}_{\rm cutoff} = 0.01$.

We remark that the equations seen in this section are valid for resting boundary with respect to the lattice. In the case of a moving boundary, it is necessary to add a momentum correction. We discuss this in the next section.

\subsection{Moving Boundaries}
\label{sec:moving-boundaries}
In the case of a moving boundary, it is necessary to introduce a moving boundary correction at wall nodes, as proposed for the HWBB in in~\citep{laddDissipativeFluctuatingHydrodynamic1991,laddNumericalSimulationsParticulate1994}. This correction term is used to generalize to moving boundaries the IBB seen in the previous paragraphs modifying each \emph{nodal} (non-wall) population in the \emph{rhs} of the equations with the following simple rule, written in terms of generalized coordinate $s$
\begin{equation}
  \text{if} \quad s < 1 \quad \text{then} \quad f(s)\to f(s)+ 2
  w_{i}\rho\frac{\boldsymbol{c}_{i}\cdot\boldsymbol{u_{w
  }}}{c_{s}^{2}}\, ,
  \label{eq:ME-solid-to-fluid}
\end{equation}
where $\bm u_w$ is the velocity of the wall at the link intersection $\bm x_W$.

\subsection{Force computation}
There exist two main ways to compute the force and torque acting on the surfaces: the stress tensor integration~\citep{inamuro_flow_2000} and the Momentum Exchange Algorithm~\citep{laddNumericalSimulationsParticulate1994}. As recommended by Mei \emph{et al.}~\citep{mei2002forceevaluation}, the momentum exchange algorithm is used for the computation of the force in the following section.

There are different numerical approaches  to compute the momentum exchange from the fluid to the surface~\citep{laddNumericalSimulationsParticulate1994,mei2002forceevaluation, hu_modified_2015,ding_extension_2003,chen_momentum-exchange_2013,li_force_2004,wen_galilean_2014} that have been summarized by Tao \emph{et al.} in~\citep{tao2016aninvestigation}. For the experiments in this paper we have decided to adopt the method described in~\citep{wen_galilean_2014}. In~\citep{wen_galilean_2014} Wen \emph{et al.} propose the following receipt to compute the force on the boundary of on lattice node
\begin{equation}
  \bm F=\left\{ \sum_{i}-\left(\left
  (\boldsymbol{c}_{i}-\boldsymbol{u}_{w}\right)f_{i}\left(x_{f}\right)
  -\left(\boldsymbol{c}_{\bar{i}}-\boldsymbol{u}_{w}\right)f_{i}^{*}\left(x_{f}\right)\right)\right\} _{\text{out}}
  \label{eq:force-boundary-node}
\end{equation}
where $\bm F$ is the force acting on the boundary surface due to one fluid lattice and the index ``out'' means that when the boundary is described as a closed surface, the force computation has to be carried out only for the ``external'' fluid.

\begin{figure*}
  \includegraphics[width=0.85\textwidth]{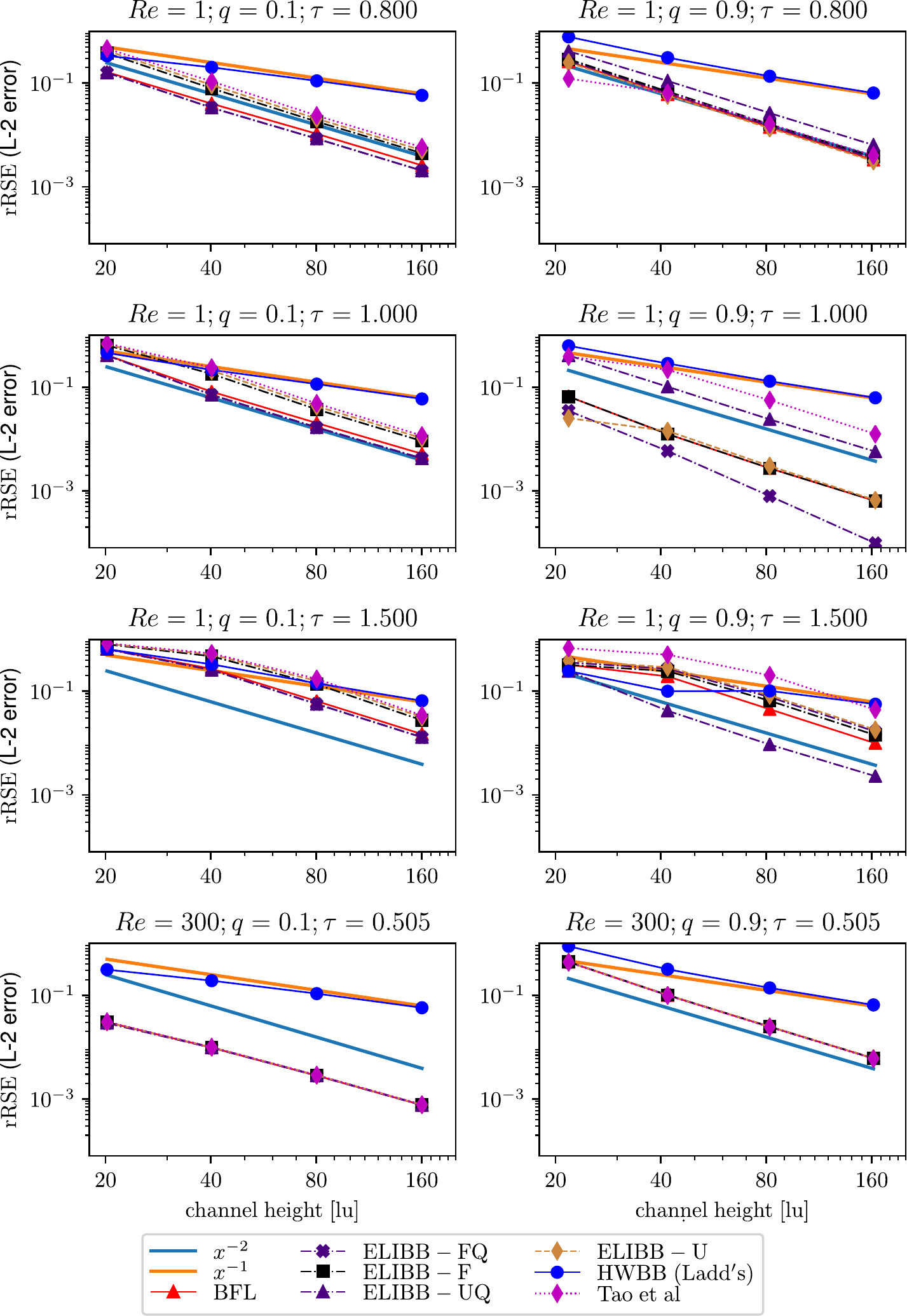}
  \caption{Convergence of rRSE for impulsively started Couette flow at
  dimensionless time $t^*=\nicefrac{\nu t}{h^2}$. Results for different Reynolds
  number $Re$, relaxation time $\tau$, and normalized distance of the walls from
  the first layer of nodes $q$. The LBM topology is D2Q9, the collision model is BGK.}
  \label{fig:couette-results}
\end{figure*}

\begin{figure}
  \centering \includegraphics[width=0.75\columnwidth]{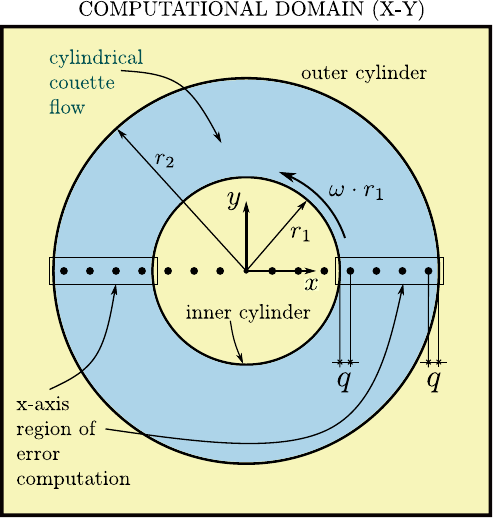}

  \caption{Representation of computational domain. In green the inter cylinders Couette flow. In the external yellow region, the fluid is at rest because the external cylinder of radius $r_{2}$ is at rest, in the inner yellow region the flow is rotation, moved by the inner cylinder of
  radius $r_{1}$ that is moving with angular velocity $\omega$. Boxed dots represent the nodes where the error along the x-axis is computed.}
  \label{fig:inter_cylinder}
\end{figure}

\begin{figure}
  \begin{centering}
    \subfloat[]
    {
    \small
    \def\svgwidth{.9\columnwidth}
  \executeiffilenewerext{img/plot0.svg}{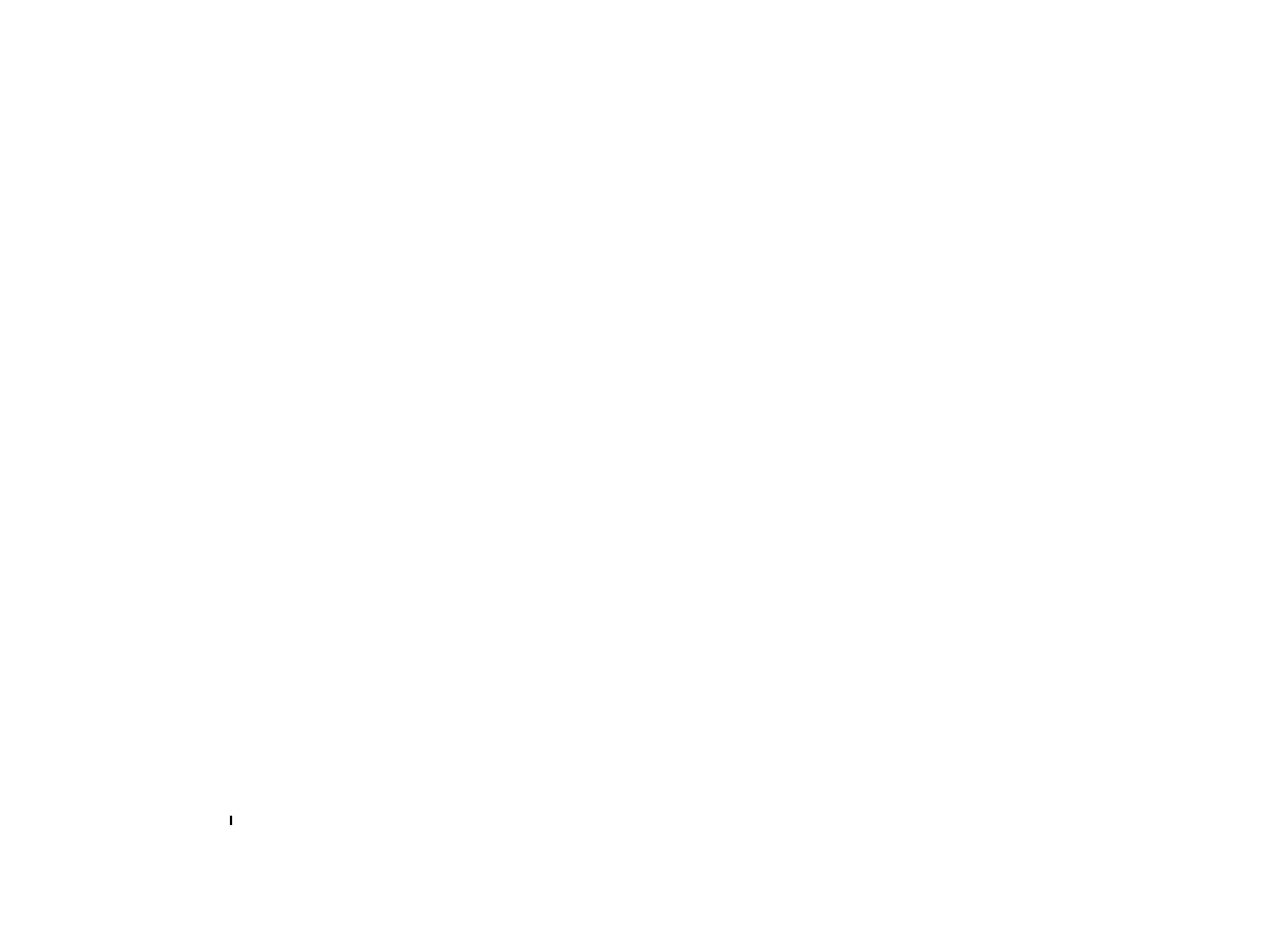}%
    {inkscape -z -D --file=img/plot0.svg --export-pdf=img/plot0.pdf --export-latex}%
  \subimport{img/}{plot0.pdf_tex_tex}%

    }

    \subfloat[]
    {
    \small
    \def\svgwidth{.9\columnwidth}
  \executeiffilenewerext{img/plot0-dLB80.svg}{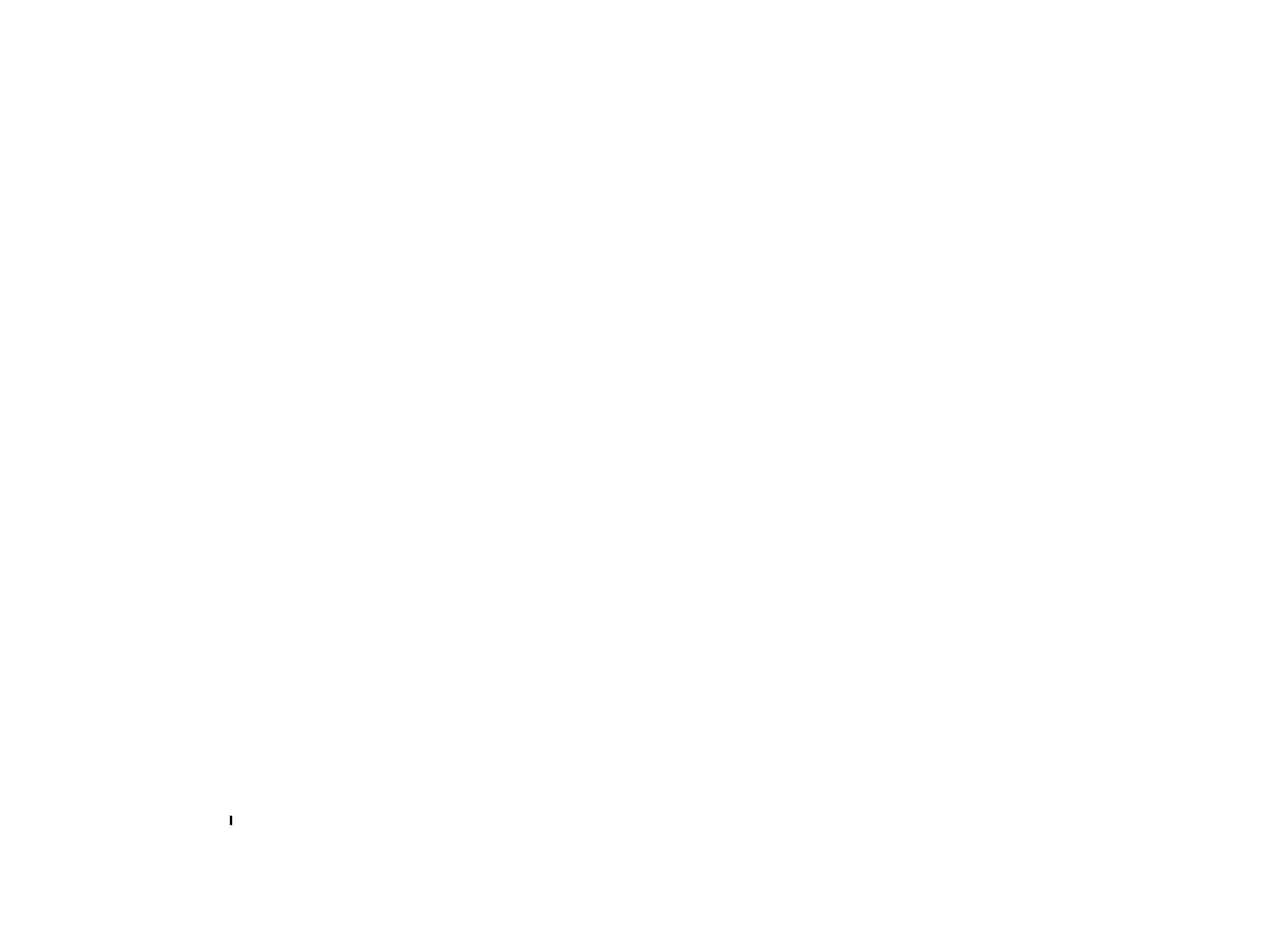}%
    {inkscape -z -D --file=img/plot0-dLB80.svg --export-pdf=img/plot0-dLB80.pdf --export-latex}%
  \subimport{img/}{plot0-dLB80.pdf_tex_tex}%

    }
  \end{centering}
  \caption{Plot of the linear error $LE=\left|\frac{u_{y}(x_{c})-u_{y}^{{\rm theory}}(x_{c})}{u_{y}^{{\rm ref}}}\right|$
  along the \emph{x-axis} for the cylindrical Couette flow characterized by $\beta=\nicefrac{1}{2}$.
  (a) $r_2-r_1=10{rm lu}$, (b) $r_2-r_1=40{rm lu}$
  The vertical lines serve as markers of the
  $x$ coordinate of intersections with the cylinders: $x<0.5$ and
  $x>2.5$ rest flow region, $0.5<x<1.0$ and $2.0<x<2.5$ cylindrical
  couette flow region, $1.0<x<2.0$ inner region (linear flow).\label{fig:LE-couette}}
\end{figure}

\section{Result and Discussion}

The variants of the novel boundary condition method have been implemented for the D2Q9 and D3Q27 lattices, in the open-source library PALABOS~\Citep{latt_palabos_2020}. In the results, we mainly used the non-equilibrium bounce-back approach to estimate the non-equilibrium component (equation~\eqref{eq:neqbb-s}) because  it is simpler to implement and gives similar results of the regularized approach (\cref{sec:hydro_limit}). Nevertheless, in the second test-case, we included a brief comparison between the results obtained with the two different methods. All the physical quantities appearing in the next sections are either in \emph{lattice units}, or in non-dimensional units specific to each problem.

\subsection{Impulsively-started unsteady Couette flow}
\label{sec:couette}
The impulsively started Couette flow is the fluid configuration obtained abruptly moving one of the two parallel walls containing a quiet fluid, from the rest position to the constant velocity $U$. In this specific case, we consider the upper wall moving along the $x$ direction and located at $y=h$ and the bottom one resting at $y=0$.

In the context of low Reynolds number flows, convective phenomena are negligible for this configuration, whose evolution is then governed by~\citep{emin_erdogan_unsteady_2002}
\begin{equation}
  \frac{\partial u}{\partial t}=v \frac{\partial^{2} u}{\partial y^{2}}
  \label{eq:ns-couette}
\end{equation}
with the following boundary and initial conditions
\begin{equation}
  \begin{array}{ll}u(0, t)=0 & \text { for all } t \\ u(h, t)=U & \text { for }
  t>0 \\ u(y, 0)=0 & \text { for } 0 \leqslant y<h\ .\end{array}
  \label{eq:bc-couette}
\end{equation}

The problem defined by equations~\eqref{eq:ns-couette} and~\eqref{eq:bc-couette}
has a solution in the form of slow converging series~\citep{emin_erdogan_unsteady_2002}
\begin{equation}
  u^*_{\rm th}=\frac{u^*_{\rm th}}{U}=\frac{y}{h}+\frac{2}{\pi} \sum_{n=1} \frac{(-1)^{n}}{n} \mathrm{e}^{-n^{2} \pi^{2} \nu t / h^{2}} \sin \frac{n \pi y}{h}\ .
  \label{eq:solution-couette}
\end{equation}
In~\citep{emin_erdogan_unsteady_2002} Erdo\v{g}an mentioned that at time $t^*=\nicefrac{\nu t}{h^2}$ equation~\eqref{eq:solution-couette} truncated at the 45th term is sufficient to obtain a numerical solution compatible with a double floating point precision numerics simulation (given that the evaluation is computationally very light, we used 100 terms in the present test case).

We used the ELIBB variants, the HWBB, the BFL, and the Tao's methods to simulate the impulsively-started Couette flow. The numerical domain is squared, bounded in the $y$ direction by the walls, and by the periodicity condition in the $x$ direction. The LBM simulation is carried out with the D2Q9 lattice and the BGK collision model. The simulations are symmetric in the two dimensions and the top and bottom layer of nodes are located at $y=q$ and $y=h-q$ (lattice units).

The \emph{relative Root Squared Error} rRSE (also known as $L_{2}$-error function), defined as
\begin{equation}
{\rm rRSE}\left(u^*\right)=\sqrt{\frac{\sum_{i}^{N}\left(u^{*}(\bm{x}_{i})-u_{{\rm th}}^{*}(\bm{x}_{i})\right)^{2}}{\sum_{i}^{N}u_{{\rm th}}^{*2}(\bm{x}_{i})}},
\label{eq:rRMSE}
\end{equation}
is used to evaluate the convergence of computational error in the cylindrical Couette flow region, where $\bm{x}_{i}$ is the coordinate of a lattice node, $u^*$ is the computed non-dimensional macroscopic velocity norm and $u^*_{{\rm th}}$ is the theoretical velocity norm given by equation~\eqref{eq:solution-couette}.

The results for the rRSE for increasing height $h$, expressed in lattice units are shown in figure~\ref{fig:couette-results}. For small relaxation numbers ($Re=330,\ \tau=5.05$) all methods show a second-order convergence and almost identical accuracy but the HWBB because $q\neq1/2$. For $\tau$ in the range $[0.8,1.5]$ the following observations can be made:
\begin{enumerate}
  \item quadratic variants of ELIBB (ELIBB-UQ, ELIBB-FQ) are generally more accurate than other methods;
  \item ELIBB-UF and ELIBB-FQ performes identically for $q<\nicefrac{1}{2}$ because
  they have the same interpolation scheme in this range;
  \item when $q\geq\nicefrac{1}{2}$ ELIBB-FQ appear to be more accurate than
  ELIBB-UQ for $\tau \lesssim 1$, but \emph{vice-versa} for $\tau \gtrsim 1$;
  \item ELIBB variants are generally more accurate than the Tao's method in this range.
\end{enumerate}

\subsection{Steady state cylindrical Couette flow}
\label{sec:couette-cy}
The cylindrical Couette flow is a common benchmark to test the accuracy of curved boundary conditions. We implemented this test case using a D3Q19 lattice. In this test case, two coaxial cylinders are placed in the center of the simulation domain. The cylinders axis is parallel to the $z$ direction, along which the periodicity condition has been imposed. The inner cylinder of radius $r_{1}$ rotates with angular velocity
\begin{equation}
  \begin{array}{ccc}
    \omega_1 = u_{\theta}/r & \text{and} & u_{\theta}=\nu\,Re/(r_2-r_{1})
  \end{array}
\end{equation}
tangential velocity, while the outer of radius $r_{2}$ is at rest. The inter-cylinder distance $r_2-r_1$ can be expressed using the cylinder ratio $\beta={r_1}/{r_2}$ parameter as $r_1 (1/\beta-1)$.

The velocity flow inside the inner cylinder is linear if the flow is laminar, with a maximum tangential velocity $u_{\theta}=1$ close to the inner cylinder and $u_{\theta}=0$ in the center. Between the two cylinders, the solution for the fluid velocity at steady state is given by the cylindrical-Couette flow tangential velocity
\begin{equation}
  u_{\theta}=\frac{\left(r_{1}^{2}-\beta^{2}r^{2}\right)\omega_{1}}{\left(1-\beta^{2}\right)r}
\end{equation}
where $r$ is the radial distance from the axis of the cylinders.
\begin{figure}
  \includegraphics[width=0.9\columnwidth]
  {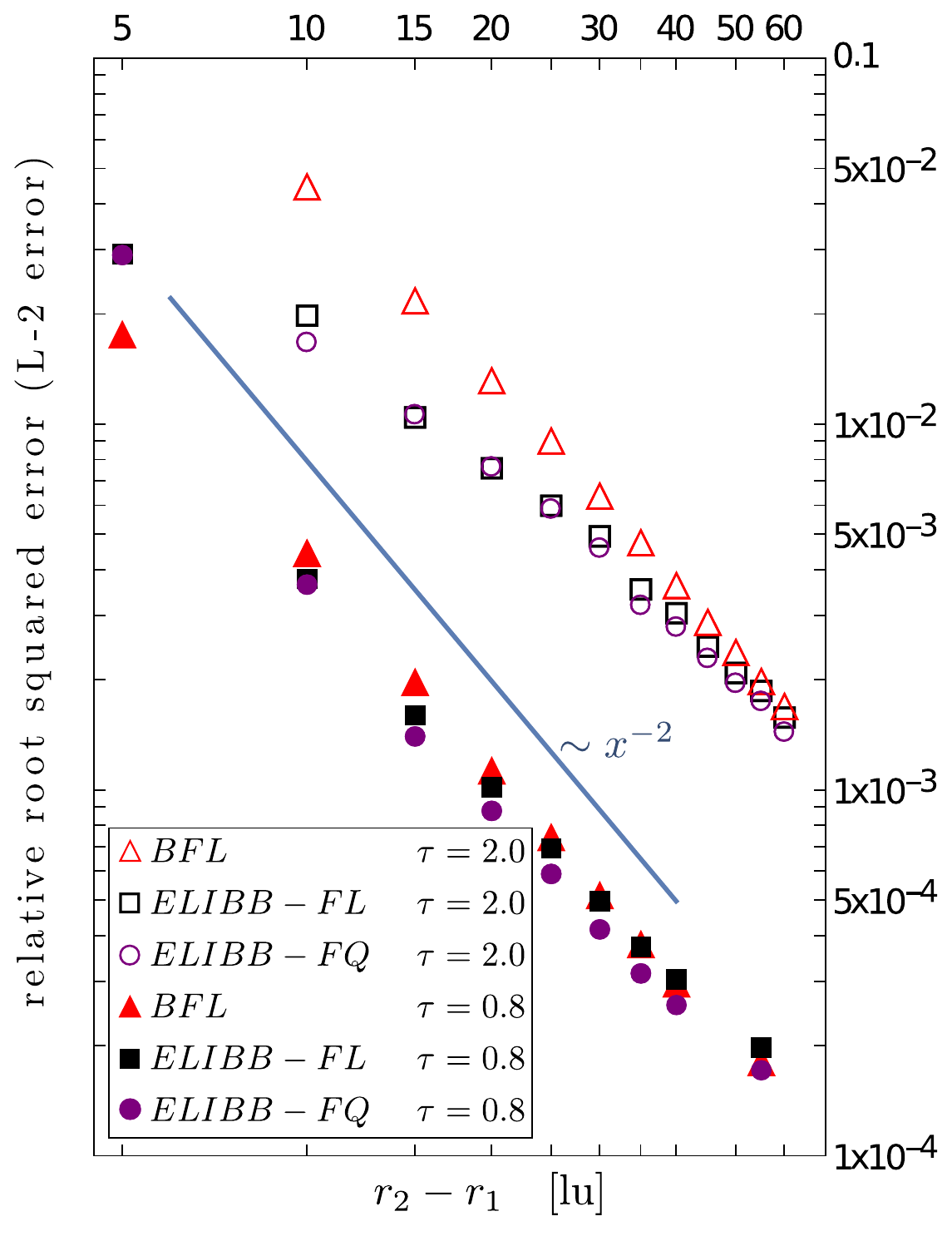}
  \caption{Cylindrical Couette flow, convergence of the \emph{relative root
  squared error} rRSE (\emph{i.e.} $L2$ error funciton)
  for the \emph{fragmented} variants of the present schemes  (ELIBB-FL,
  ELIBB-FQ) and for the Bouzidi \emph{et al.} (BFL) method.
  Parameters: TRT collision model with $\Lambda=\nicefrac{3}{16}$,
  $\beta=0.5$.\label{fig:convergence-fragmented}
  }
\end{figure}
\begin{figure}
  \includegraphics[width=0.9\columnwidth]
  {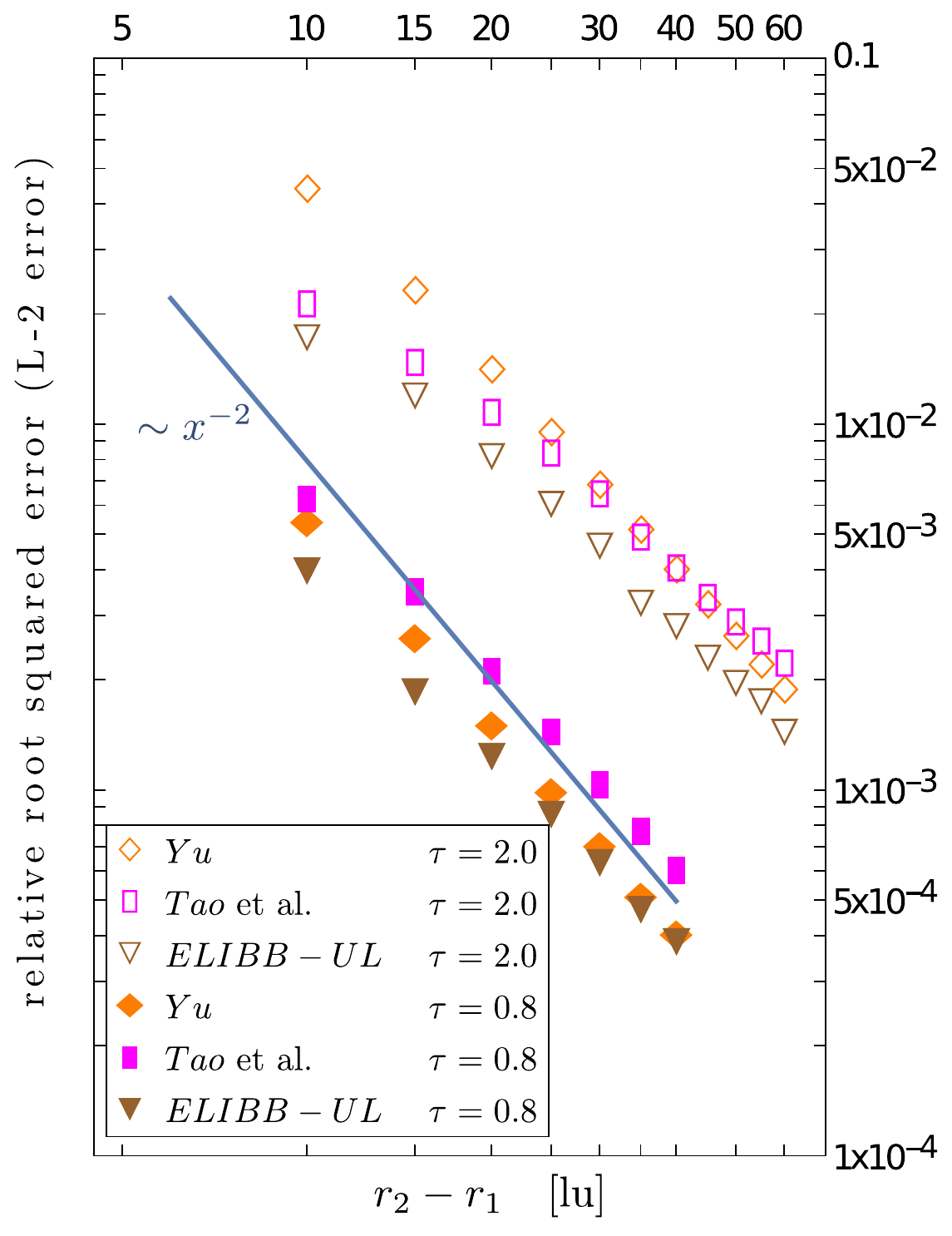}
  \caption{Cylindrical Couette flow, convergence of the \emph{relative root
  squared error} rRSE (\emph{i.e.} $L2$ error function)
  for the \emph{unified} variant of the present schemes  (ELIBB-UL)
  and for the Yu \emph{et al.} method.
  Parameters: TRT collision model with $\Lambda=\nicefrac{3}{16}$,
  $\beta=0.5$.\label{fig:convergence-unified}
  }
\end{figure}

The \emph{relative Root Squared Error} rRSE, defined by equation~\eqref{eq:rRMSE} is used to evaluate the convergence of computational error in the cylindrical Couette flow region (see figure \ref{fig:inter_cylinder}), wherein this case $u^*_{{\rm th}}=u_\theta$ is the theoretical velocity norm. The rRSE is also used to evaluate the error of the tangential velocity along the \emph{x-axis,} in this case, the expression reads\begin{equation}
{\rm rRSE}\left(u_{\theta}(y_{{\rm c}})\right)=\sqrt{\frac{\sum_{i}^{N}\left(u_{\theta}(x_{i},y_{{\rm c}})-u_{{\rm th}}(x_{i},y_{{\rm c}})\right)^{2}}{\sum_{i}^{N}u_{{\rm th}}^{2}(x_{i},y_{{\rm c}})}},\label{eq:rRMSErotatingCylinders}
\end{equation}
where $y_{c}$ is the $y-$coordinate of the cylinders axis and the index $i$ refers to the index of the node along the $x$-direction. To visualize the numerical error along the $x$-axis, we also use the linear error (LE), that we define as:
\begin{equation}
  LE\left(x_{i}\right)=\left|\frac{u_{y}(x_{i},y_{c})-u_{y}^{{\rm th}}(x_{i},y_{c})}{u_{y}^{{\rm th}}}\right|
  \label{eq:LE}
  \, ,
\end{equation}
where $x_i$ is the $x$ coordinate of a node located at $y_c$, which is the $y$ coordinate of the center of the cylinders.
\subsubsection{Space distribution of the linear error}
figure~\ref{fig:LE-couette} show the trend of $LE$ for two resolutions $r_2-r_1$ ($10\, {\textrm{lu}}$ and $40\, {\textrm{lu}}$) and for seven different boundary conditions: the BFL method~\citep{bouzidi2001momentum} (linear version), the unified scheme of Yu \emph{et al.}~\citep{yuaunified}, the single node method of Tao \emph{et al.}~\citep{tao2018onepoint} and the three variants of the ELIBB, plus the HWBB in the 10lu case. The simulated experiments were carried out for $Re=10$ in the case of the TRT collision model using a $D3Q27$ lattice layout and for $q=0$ (see figure~\ref{fig:inter_cylinder}). From these figures is possible to see that,
\begin{enumerate}
  \item fragmented methods BFL, ELIBB-FL and ELIBB-FQ generally more accurate than the unified methods ELIBB-U and Tao;
  \item the space error distribution is similar for all the methods;
  \item the only local methods that performs similarly to the BFL are the
  ELIBB-F and ELIBB-FQ;
  \item for low resolutions the present \emph{fragmented} schemes (ELIBB-F and ELIBB-FQ)
  show smaller error than the BFL.

\end{enumerate}

\begin{figure}
  \centering
  \includegraphics[width=0.45\textwidth]{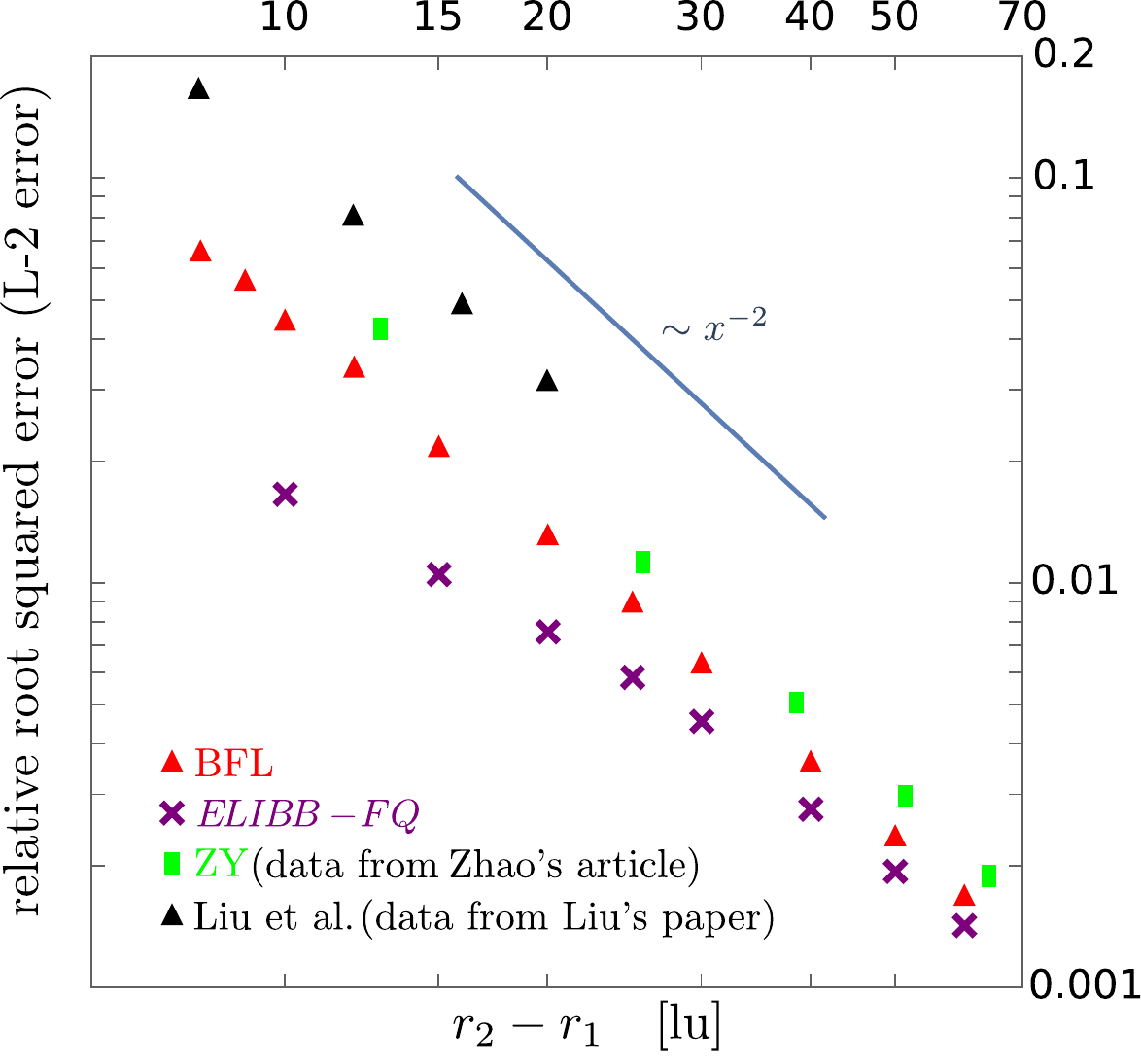}
  \caption{Comparison of the rRSE (L-2 error) between the BFL and ELIBB-Q
  methods simulated for this paper and other results from the literature.
  The error is relative to the cylindrical Couette region (blue region in
  figure~\ref{fig:inter_cylinder}).
  For the
  results of the Liu \emph{et al.} method data from~\citep{liu2019lattice} is
  used. For the ZY methods data from~\citep{zhao2017singlenode} is used.
  Other parameters: TRT $\Lambda = \nicefrac{3}{16}$, lattice topology for the cases simulated = D3Q27,
  $\tau = 2$.\label{fig:comparison-liuZY}}

\end{figure}

\subsubsection{rRSE convergence}
All the previously mentioned methods show approximately second order convergence as displayed in figures~\ref{fig:convergence-fragmented} and \ref{fig:convergence-unified}. The results appearing in the figures concern a low reynolds fluid flow $Re = 10$ between cylinders with diameters ratio $\beta = \frac{r_a}{r_b} \nicefrac{1}{2} $. The choosen collision model is the TRT with magic parameter $\Lambda =\nicefrac{3}{16}$, viscous relaxation numbers $\tau = 0.8$ and $\tau = 2$. Those figures also confirm the observations discussed above: the three \emph{fragmented} methods tested (BFL, ELIBB-FL, ELIBB-FQ) are generally slightly more accurate than the three \emph{unified} techniques (Yu, Tao, ELIBB-UL).

We finally compare the results of the ELIBB-FQ method with some results from the literature, namely the Liu \emph{et al.}~\citep{liu2019lattice} method and the ZY method \citep{zhao2017singlenode}. The Liu \emph{et al.} can be seen as an extension of the Tao \emph{et al.}, while in the ZY method the \emph{single-node} characteristic is achieved setting $f(\fMinusQ,t+1)\approx f(\fMinusQ,t)$. Even if figure~\ref{fig:comparison-liuZY} cannot be considered a direct comparison, it suggests that the ELIBB-FQ method results in a more accurate solution for the considered set of parameters. Note that in the steady-state test described by figure~\ref{fig:comparison-liuZY} the ZY and BFL method should deliver the same results given that the ZY method is a first-order in time approximation of the BFL. Therefore, the different results between the green and the red points in figure~\ref{fig:comparison-liuZY} are most probably caused by the different lattice topology, error computation, numerical precision, or position relative to the lattice of the cylinders axis.

\begin{figure*}

  \includegraphics[scale=0.8]
  {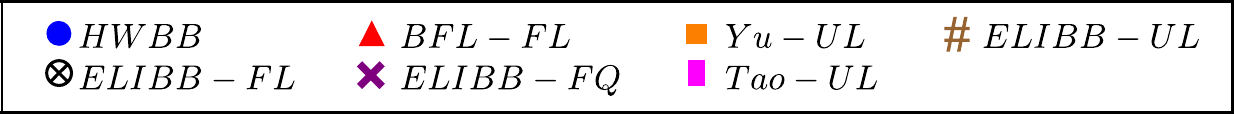}
  \centering \subfloat[]{
  \includegraphics[viewport=0bp 0bp 359bp 227bp,width=0.9\columnwidth]{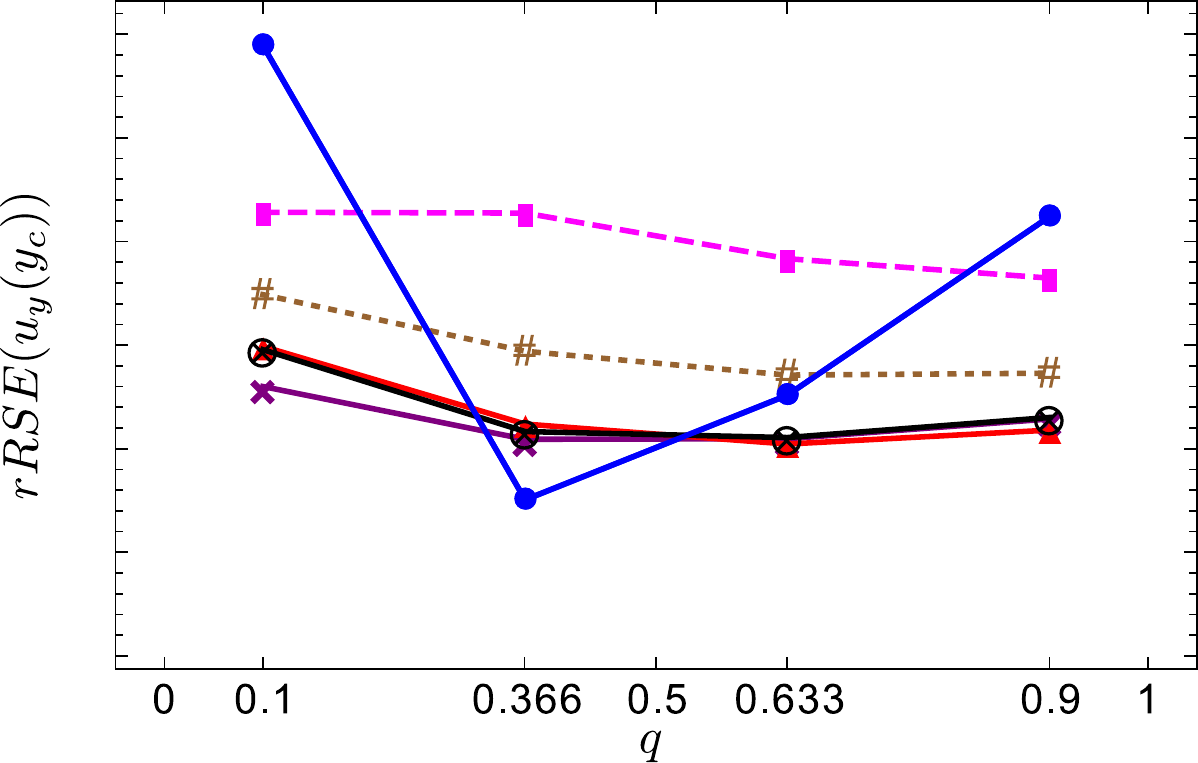}

  }\subfloat[]{\hspace{-0mm}
  \includegraphics[viewport=0bp 0bp 359bp 227bp,width=0.9\columnwidth]{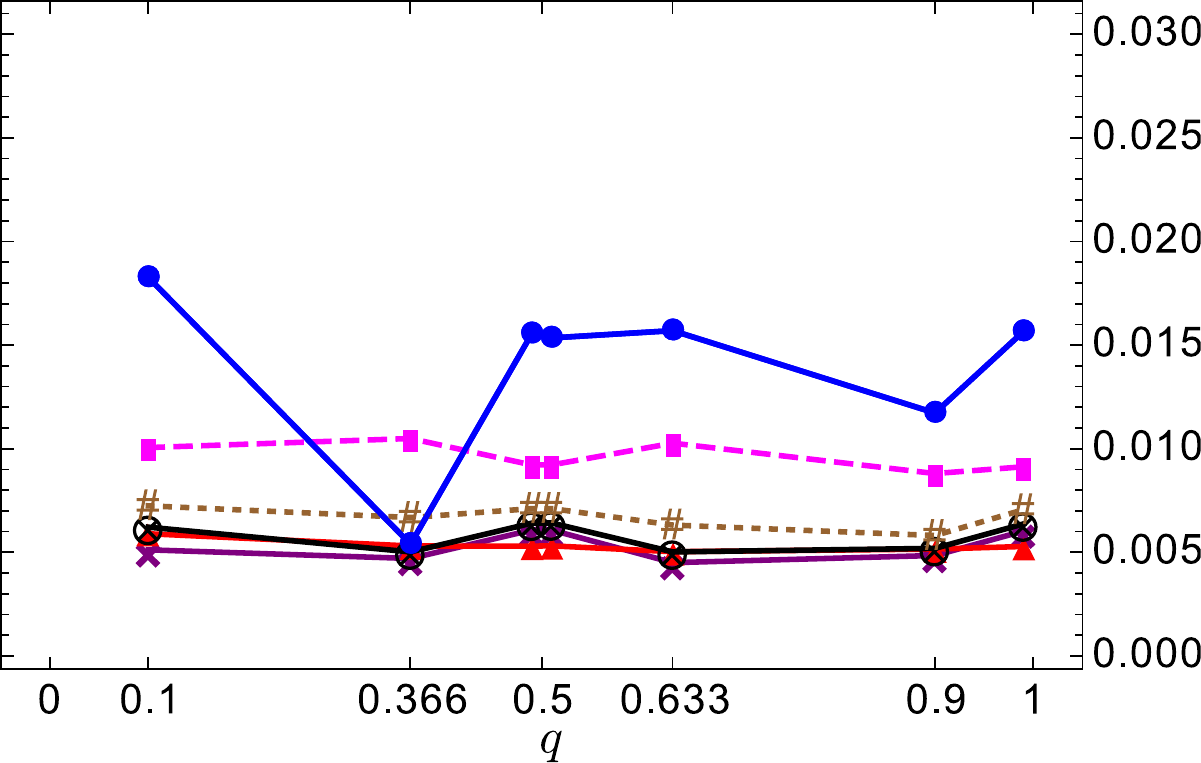}
  }
  \caption{Comparison of rRSE (\ref{eq:rRMSErotatingCylinders}) for two
  different resolutions (a) $r_{2}-r_{1}=20+2q$ , (b) $r_{2}-r_{1}=30+2q$
  as a function of $q$ for a set of methods. Parameters used for the simulation:
  $Re=10$, TRT collision model, $\Lambda=\nicefrac{3}{16}$, $\tau_{\nu}=1$.
  \label{fig:Comparison-q}}
\end{figure*}

\begin{figure*}
  \centering
  \includegraphics[scale=0.8]{img/fig9.pdf}

  \subfloat[]{
  \includegraphics[scale=0.50]{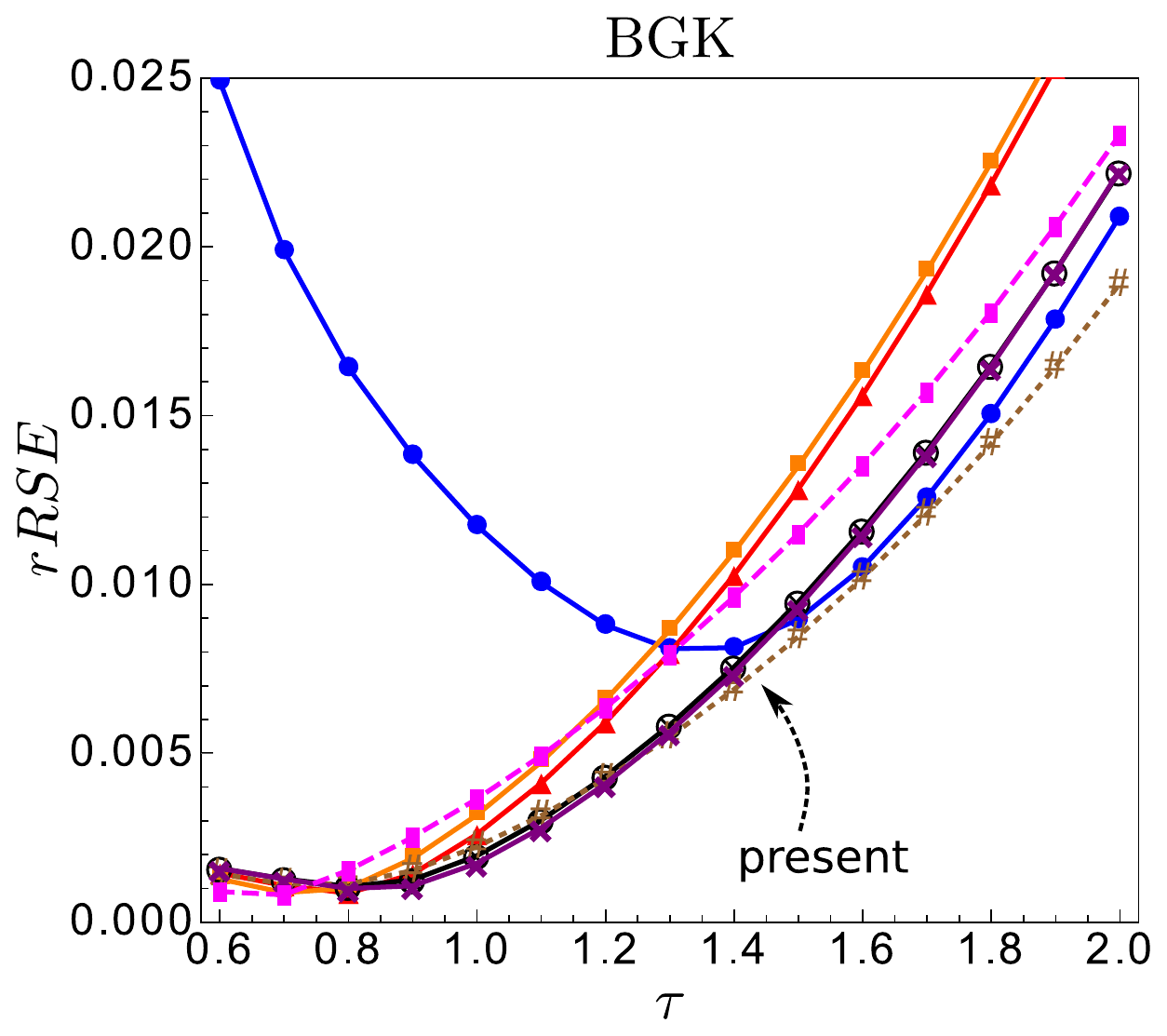}\label{fig:tau_dependence-bgk}
  }
  \subfloat[]{\hspace{-1.2cm}
  \includegraphics[scale=0.50]{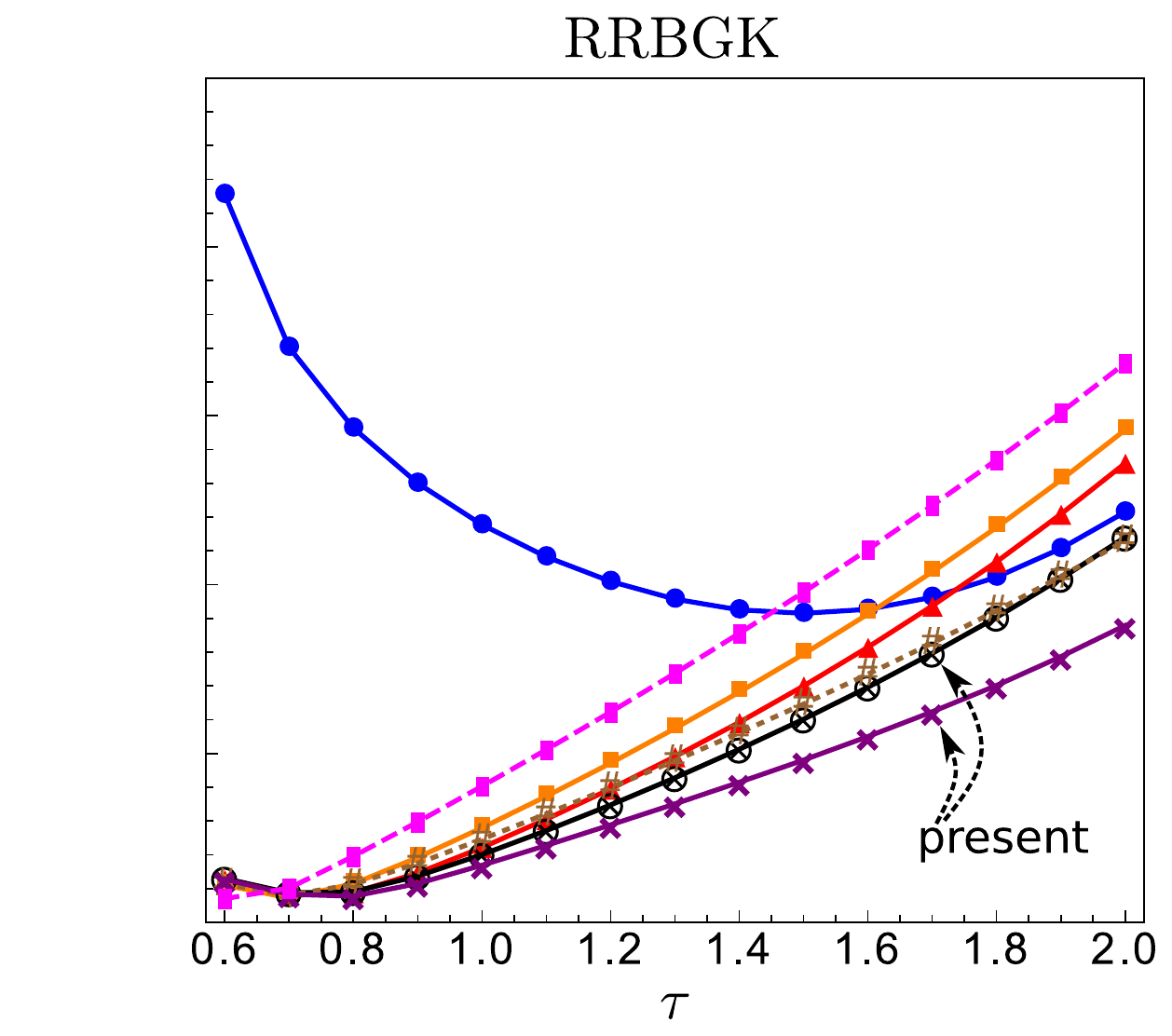}
  }
  \subfloat[]{\hspace{-1.2cm}
  \includegraphics[scale=0.50]{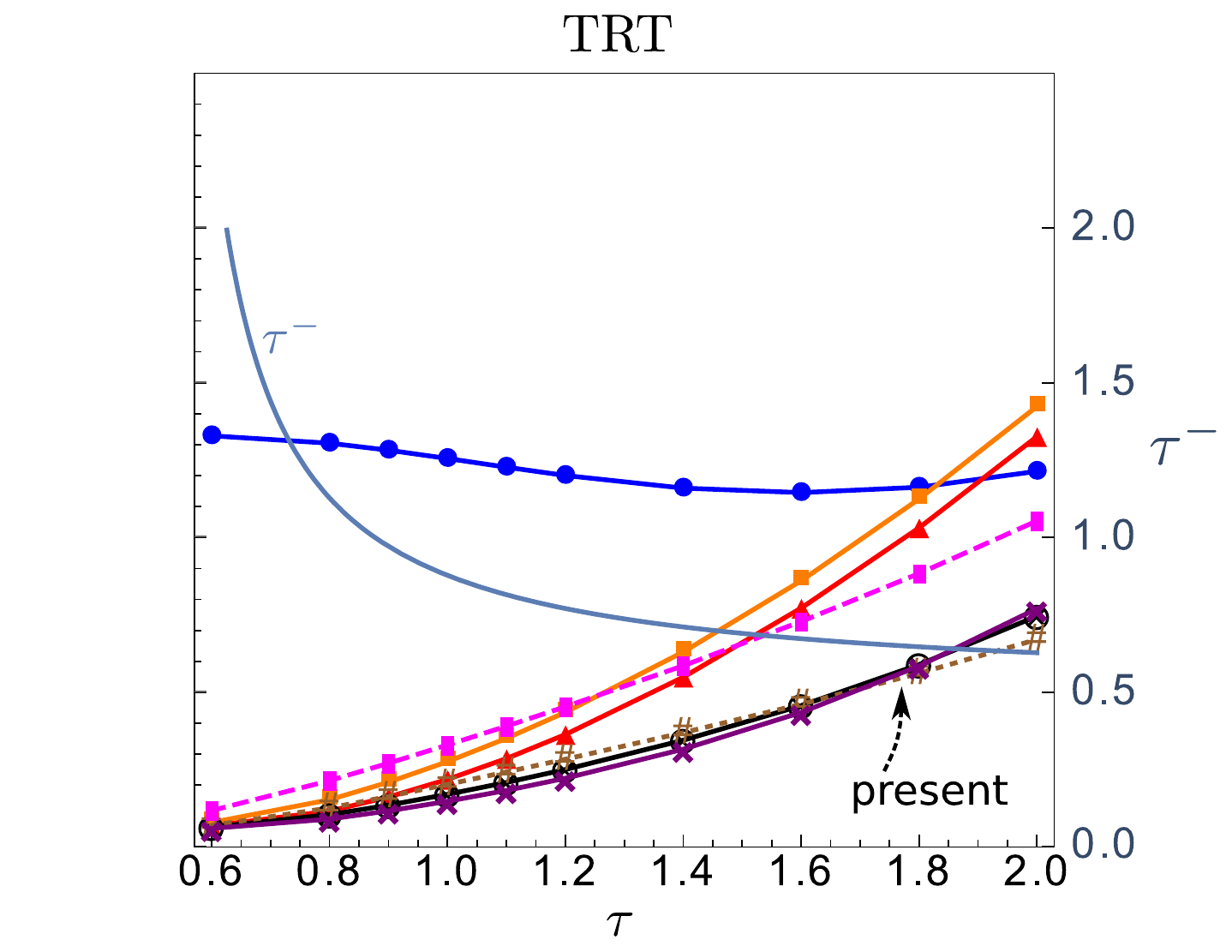}
  }
  \caption{Sensitivity to $\tau$ of the relative root squared error
  rRSE~\eqref{eq:rRMSErotatingCylinders}, \emph{i.e.} L2 error function, in
  the case of cylindrical Couette flow. Parameters of the simulation:
  $Re=10$, inter-cylinder distance $L=20\,{\rm lu}$, $\beta = 0.5$. For the
  TRT collision model and the magic parameter $\Lambda$ is fixed to $3/16$.
  \label{fig:tau_dependence}}

\end{figure*}

\subsubsection{rRSE as a function of $q$}
Some authors~\citep{chen2014acomparative} suggested that unified methods can help to improve the stability of moving boundaries because they do not need to change interpolation points crossing the midpoint between two nodes. We tried to assess this behavior, modifying the parameter $q$ for different resolutions keeping the other parameters constant. The results are shown in figure~\ref{fig:Comparison-q}. From the results of this specific experiment, it is not possible to confirm this insight, since the distance $q$ has a small impact on error levels for all methods but the HWBB. This kind of test has also been recently carried out for BFL and Yu methods in~\citep{zhangvelocity}, for a Poiseuille flow with moving walls.

\subsubsection{Non-equilibrium computation at the wall}
\begin{figure}
  \includegraphics[width=0.9\columnwidth]{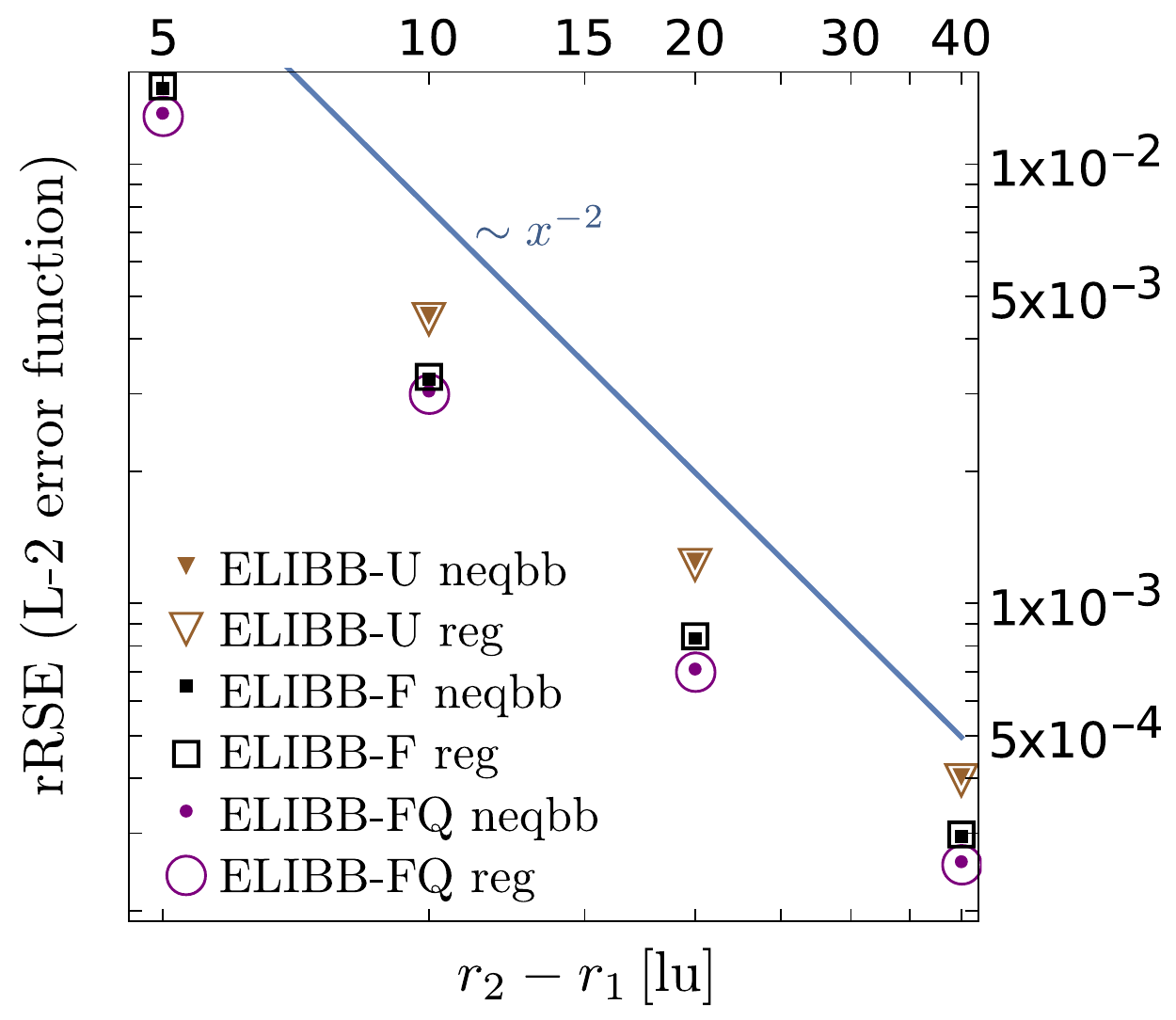}
  \caption{Comparison of the convergence of the rRSE for
  the proposed ELIBB methods variants in the case of the two different
  non-equilibrium component computation. The \emph{neqbb} acronym stands for
  method described by equations~\eqref{eq:neqbb-s} and the acronym \emph{reg}
  stands for the method described by equation~\eqref{eq:fi_neq}).
  }
  \label{fig:neq0vsneq1}
\end{figure}

Section~\ref{sec:approx-wall-pop} provided two ways of computing the non-equilibrium component at the wall position with a first-order approximation. In our experience, the two methods lead to similar results for low Re numbers. In figure~\ref{fig:neq0vsneq1}, we present an example of a comparison of the two methods. The figure shows that the technique used for non-equilibrium computation does not seem to have any impact on the error, at least for this range of parameters. However, there is evidence that regularized approaches are more stable for higher values of the Reynolds number. Hence, as future work, it will be interesting to further compare both approaches in the low viscosity regime.

It is somehow important to mention that the non-equilibrium bounce-back (equation~\eqref{eq:neqbb-s}) is generally easier to implement than the \emph{regularized} approach, because the latter can be used to impose both macroscopic values and their gradient similarly to Robin boundary conditions. This can be useful for advanced boundary modeling, like for turbulent wall modeling~\citep{malaspinas_wall_2014}.

\subsubsection{Evaluation of the viscosity dependence of the error}
The sensitivity of the results to the viscosity relaxation time $\tau$, under diffusive scaling, has been investigated for BGK, TRT, and RRBGK collision models. In the case of BGK (figure~\ref{fig:tau_dependence-bgk}) all models, and especially the HWBB, in accordance with the work by Ginzburg \emph{et al.}~\citep{ginzburg2003multireflection,krugeretal.2016thelattice} For the lower values of $\tau$, the beneficial displacement of the boundary caused by the viscosity dependence is higher than the accuracy deterioration owed to the increased time-step (consequent to increased $\tau$). For higher values of $\tau$, the error rise for all methods. This confirms a non-negligible viscosity dependence effect in these methods if coupled with the simple BGK. It is anyhow interesting to notice how, in this experiment and for higher values of $\tau$, the simple half-way bounce-back method performs better than any interpolated version in the BGK case. In the case of the TRT collision model, the $\tau$ dependence of the HWBB is almost wiped out and its rRSE becomes almost a horizontal line in figure~\ref{fig:tau_dependence}. As a final comment on figure~\ref{fig:tau_dependence}, one may notice that the present ELIBB schemes show a resilient behavior for high values of $\tau$: this fact combined with the good accuracy at low resolutions make the ELIBB particularly robust methods for coarse space and time resolution simulations.

\subsubsection{Mass conservation violation}
\begin{figure}
  \includegraphics[scale=0.6]
  {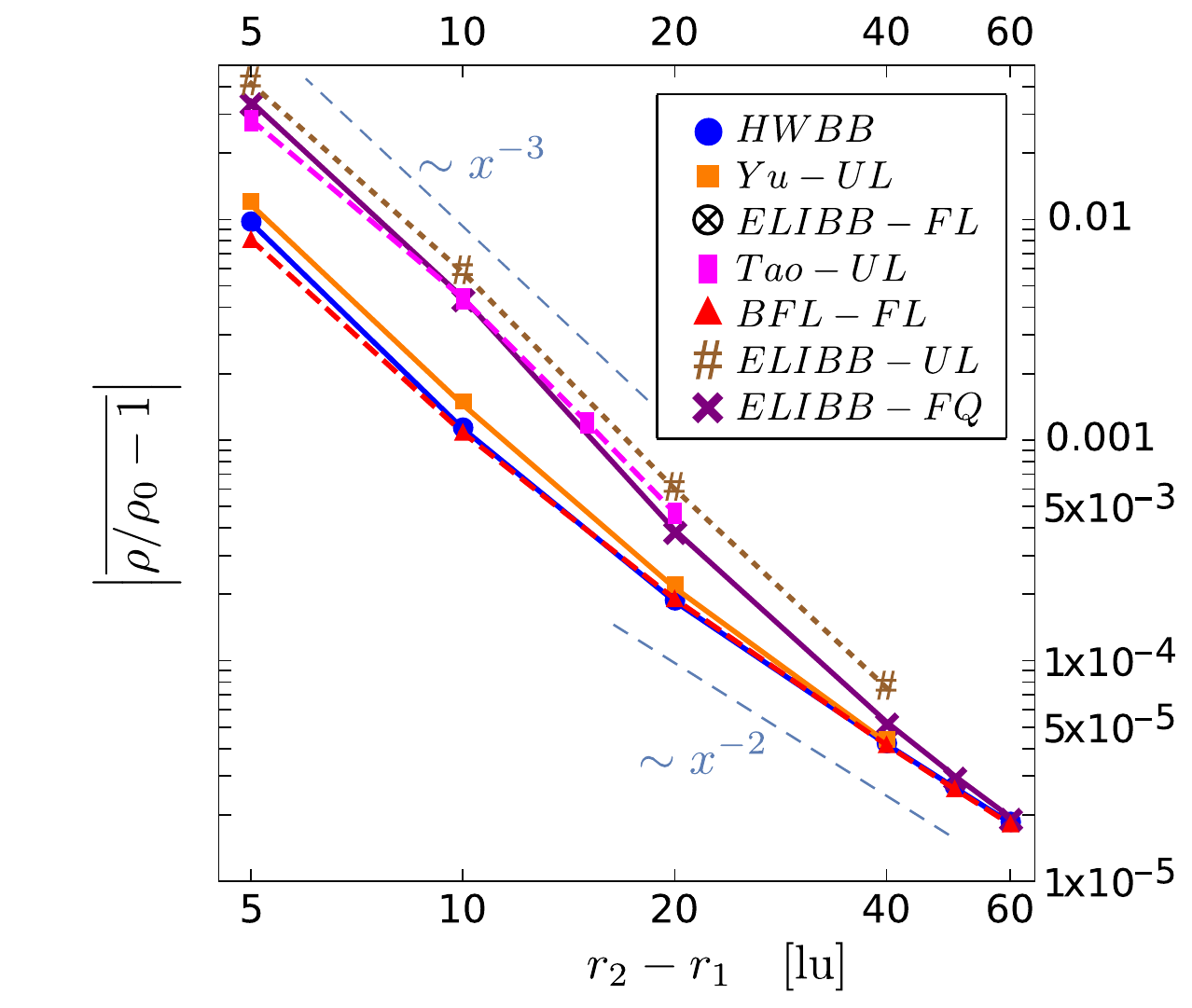}

  \caption{Average density fluctuation $|\rho_{LB}-1|$ in the cylindrical
  Couette flow region at $t^* = t/t_{ref} = 3.0$ where $t_{ref}=(r_2-r_1)/u$.
  Parameters: TRT collision model $\Lambda =
  \nicefrac{3}{16}$, $\tau = 1$.\label{fig:Comparison-mass-tau1}}
\end{figure}

\begin{figure}

  \includegraphics[scale=0.6]
  {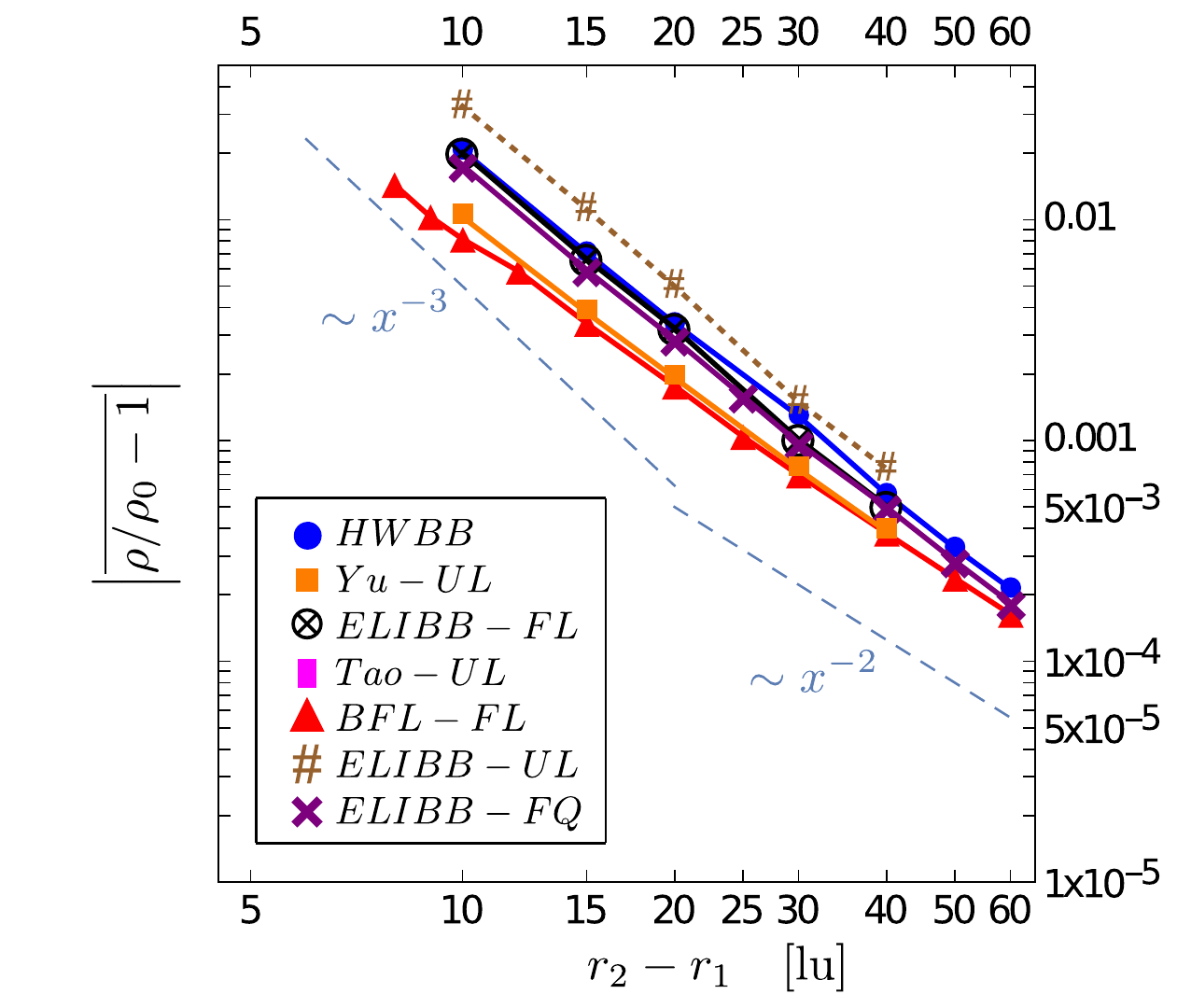}

  \caption{Average density fluctuation $|\rho_{LB}-1|$ in the cylindrical
  Couette flow region at at $t^* = t/t_{ref} = 3.0$ where $t_{ref}=(r_2-r_1)/u$.
  Parameters: TRT collision model $\Lambda =
  \nicefrac{3}{16}$, $\tau = 2$.
  \label{fig:Comparison-mass-tau2}}
\end{figure}

A common issue with interpolated bounce-back is the violation of mass conservation. We investigated this concern computing the average density fluctuation $\left|\sum_i^N(\rho_i-\rho_0)/N\right|$ in the Taylor-Couette region at non-dimensional time $t^* = t/t_{ref} = 3.0$ where $t_{ref}=(r_2-r_1)/u$ and $N$ is the number of nodes in the Taylor-Couette region; the results are presented in figures~\ref{fig:Comparison-mass-tau1} and~\ref{fig:Comparison-mass-tau2}. The conclusion is that, even though the ELIBB are pretty accurate at low time and space resolutions, in this regime they show a higher mass violation. Nonetheless, for finer meshes  they exhibit similar values of the average density fluctuation.


\begin{figure}
  \includegraphics[width=0.45\textwidth]{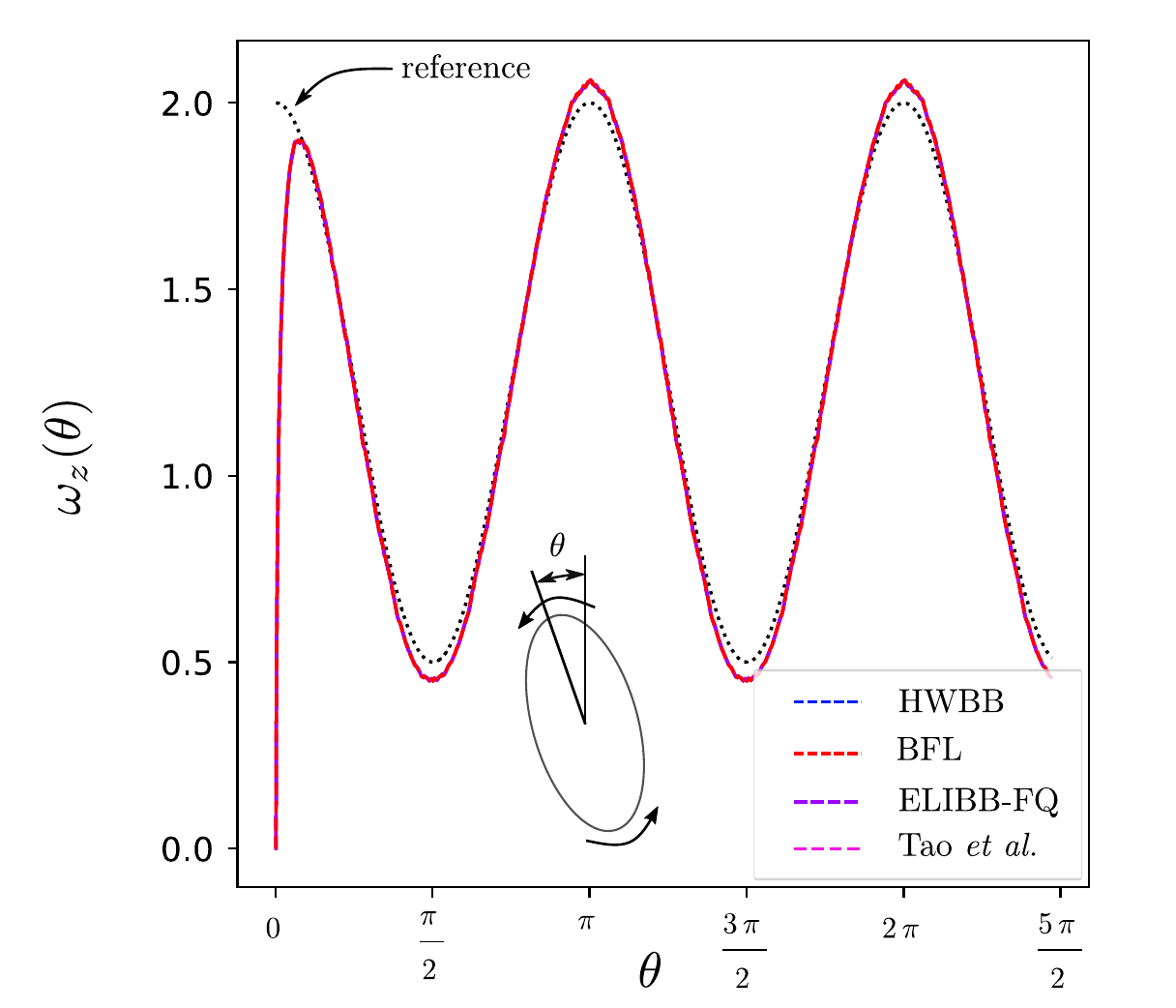}
  \label{fig:omega3d-3d}
  \caption{Jeffery's orbit described by a 3D prolate ellipsoid
  characterized by a ratio of  radii $\beta =\nicefrac{r_a}{r_b}= 2$.
  Plots
  of the angular velocity
  obtained with different methods compared with the analytical solution.
  Parameters of the simulations $\tau = 2$, \emph{BGK} collision model,
  $Re=4$, $H=10 r_e$ (channel
  height),
  $r_e=\sqrt{r_a^2+2\cdot r_b^2} = 30$ (equivalent radius).\label{fig:ang-vel}}
\end{figure}
\subsection{Jeffery's orbit: Ellipsoidal Cylinder and Ellipsoid Rotation}
\label{sec:jeffery}
Jeffery's orbit is a common benchmark test for curved boundary conditions. It describes the rotation of ellipsoidal objects induced by a shear flow in Stoke's regime~\citep{jeffery1922themotion}. Therefore, it is well suited to verify the capability of a numerical method to describe a fluid-solid interaction problem. In our experiment, the ellipsoid is located at the center of a channel. The channel is delimited in the $y$ direction by two horizontal walls, that are impulsively moved along the $x$ direction at the beginning of the simulation with the fluid at rest. In the initial condition, the prolated ellipsoid lies at the center of the channel, in vertical position, with its longer diameter aligned with the $y$ direction. The computational domain is periodic in the $x$ and $z$ directions. At the time $t=0$ the upper wall is abruptly accelerated to its terminal velocity $u_{\textrm{lid}}$, such that it generates the following strain rate in the channel
\begin{equation}
  \dot \gamma = \frac{u_{\textrm{lid}}}{H}
\end{equation}
where $H$ is the channel height.
The ellipsoid start accelerating until it reaches its steady state Jeffery's
orbit, that reads~\citep{jeffery1922themotion}
\begin{equation}
  \dot{\theta}=\frac{\dot{\gamma}}{r_{e}^{2}+1}\left(r_{e}^{2} \cos ^{2} \theta+\sin ^{2} \theta\right)
\end{equation}
where $\theta$ is the inclination of the ellipsoid axis corresponding to the major radius $r_a$ with respect to the $y$-axis (the vertical one), $\dot\theta$ is the corresponding angular velocity and $r_e=\sqrt{r_a^2+2\cdot r_b^2}$ is the equivalent radius. The Reynolds number in this scenario is redefined with the shear stress
\begin{equation}
  \mathrm{Re} = \frac{\dot \gamma r^2}{\nu}= \frac{u_{lid} \nicefrac{r_e^2}{H}}{\nu}\, .
\end{equation}

In this experiment, we describe the ellipsoid with a thin-shell surface whose dynamics are computed following the rigid body motion equations. The thin-shell representing the ellipsoid is filled with fluid. Nevertheless, the internal fluid is virtual and has no impact on the dynamics. This because the forces from the fluid to the rigid body are computed only considering the external fluid, using equation~\eqref{eq:force-boundary-node}. All the investigated methods lead to similar values of the ellipsoid angular velocity evolution in time in the case of a prolate ellipsoid with a diameter ratio $\beta=\nicefrac{r_a}{r_b}=2$. In particular in figure~\ref{fig:ang-vel}, we compared the results of a 3D ellipsoid angular velocity evolution for the HWBB, BFL, and the present ELIBB-FQ, but the results are similar for all the ELIBB variants and the other methods investigated in the previous section.

\begin{figure}[t!]
  \centering
  \includegraphics[width=0.45\textwidth]{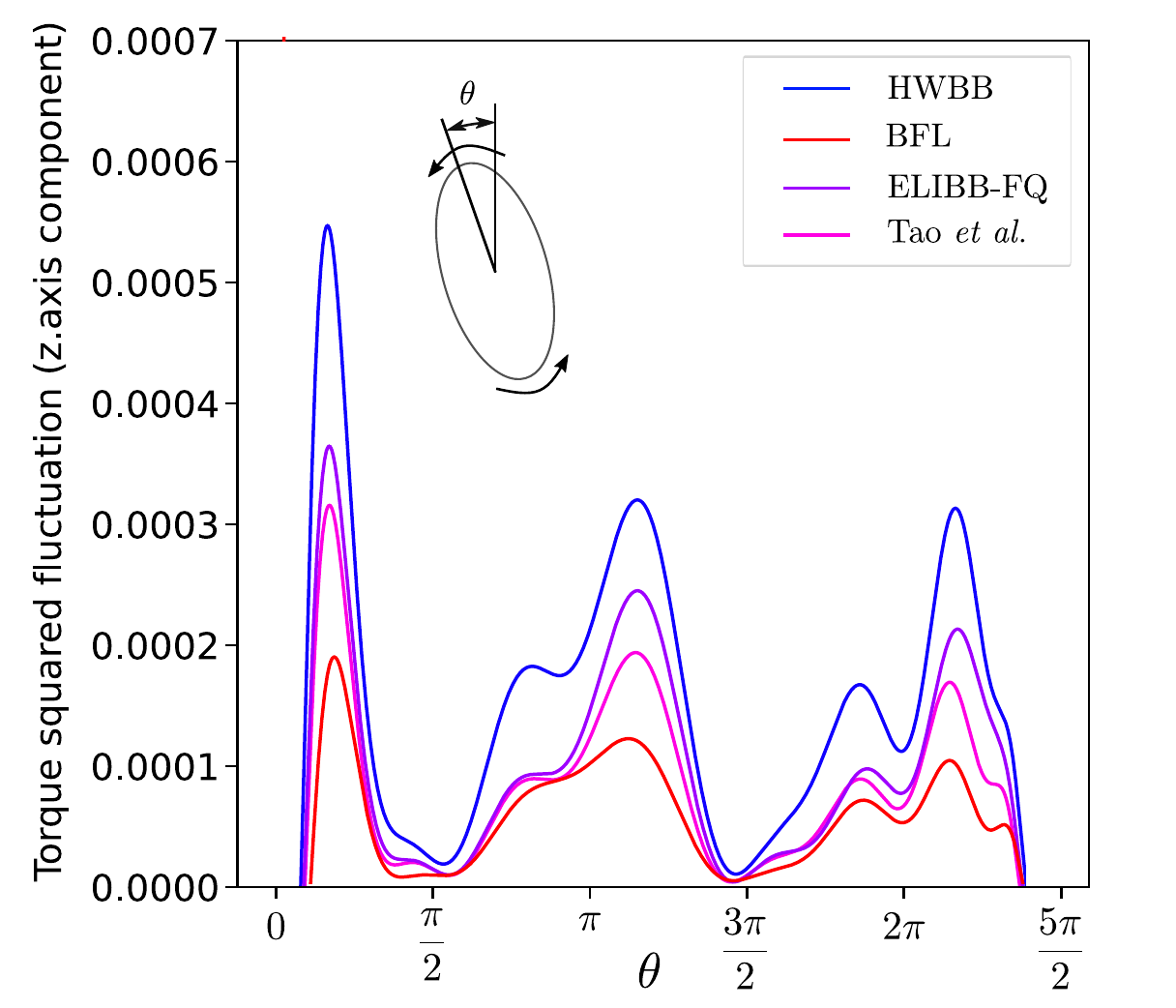}

  \caption{Jeffery's orbit described by a 3D prolate ellipsoid
  characterized by a ratio of  radii $\beta =\nicefrac{r_a}{r_b}= 2$.
  Plots of the
  squared fluctuation of the torque obtained with equation~\eqref{eq:TSF30}. \label{fig:torque-fluct-3d-nolir}}
\end{figure}
\begin{figure}[t!]
  \centering
  \includegraphics[width=0.45\textwidth]{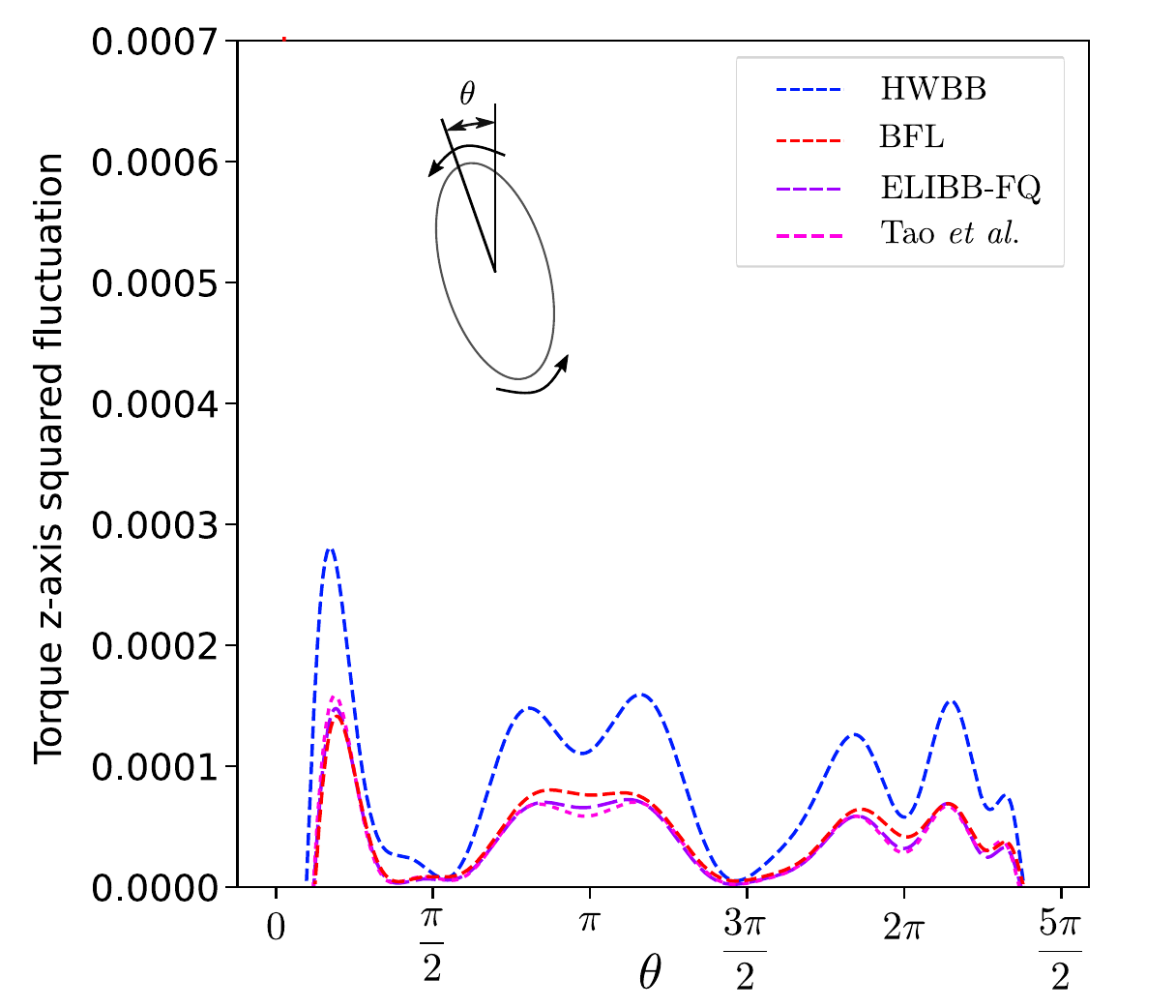}
  \label{fig:torque-fluct-3d}
  \caption{Jeffery's orbit described by a 3D prolate ellipsoid
  characterized by a ratio of  radii $\lambda =\nicefrac{r_a}{r_b}= 2$.
  Plots of the
  squared fluctuation of the torque obtained with equation~\eqref{eq:TSF30}.\label{fig:torque-fluct-3d-lir}}
\end{figure}

\subsubsection{Numerical noise}
When dealing with moving boundaries, LBM boundary methods give rise to spurious pressure oscillations. One of the main sources of these oscillations is the results of some nodes, named \emph{fresh nodes}, changing the side of the boundary surface. In the upwind part of the surface \emph{fresh nodes} appear as uninitialized nodes with wrong populations' values: this causes the triggering of pressure waves. The process of recomputing the values for the distribution functions in the \emph{fresh nodes} is called \emph{refilling} and can reduce the magnitude of pressure oscillations. The \emph{refill} methods have recently been compared in~\citep{tao2016aninvestigation}. The authors indicate the \emph{local iteration refill} (LIR)~\citep{chen2014acomparative} as the most effective in reducing oscillations. For this reason, we choose the LIR to correct the spurious pressure oscillations, implementing it in a slightly modified version to make it consistent for the case of thin-shell boundaries (see \cref{sec:LIR}). In our experiments, using a thin-shell approach, the simulations are stable even if no refilling algorithm is used. Therefore, we tested different boundary conditions before and after the implementation of the LIR.

The pressure oscillations due to the boundary motion are transferred to the rigid body through the momentum coupling, leading to a noisy torque time evolution. The measure of the oscillations in the torque (or resulting force) acting on the rigid body, is a common way to estimate the magnitude of pressure oscillations~\citep{caiazzo_boundary_2008,liAnalysisAccuracyPressure2016,tao2016aninvestigation}. It is anyhow important to notice that perturbations in the torque acting on the body are only an indirect measure of the effect of pressure waves and can be also influenced by the techniques used for the force computation.

We decided to use a qualitative approach to compute the TSF that allows producing smooth graphs that are easy to compare. The detail of the computations is shown in \cref{sec:qualitative_approach}. Here, we only point out that the presented values of the TSF are interpolated values using best fitting polynomials. Therefore, they should be interpreted as \emph{qualitative} measures that do not aim at accurate measurements.

Using the momentum exchange algorithm and without refilling techniques, local methods show higher fluctuations (figure~\ref{fig:torque-fluct-3d-nolir}). Nevertheless, after the implementation of the LIR, both local and non-local interpolated methods show similar performance in terms of torque squared fluctuation (figure~\ref{fig:torque-fluct-3d-lir}).

\begin{figure}
  \centering
  \includegraphics[width=0.45\textwidth]{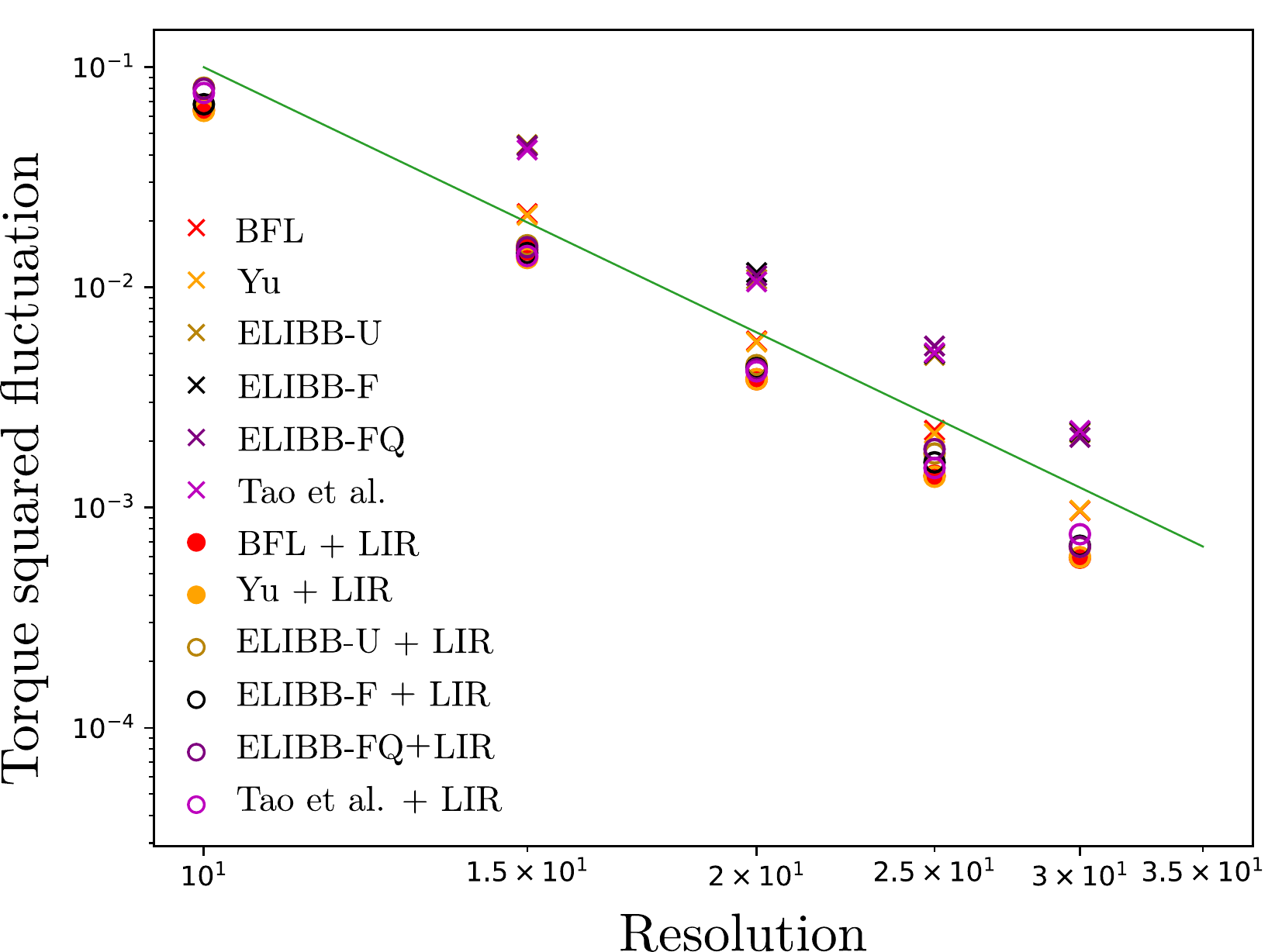}

  \caption{Average root squared fluctuation of the torque
  for an ellipsoidal cylinder (with ratio between axes $\nicefrac{d_a}{d_b}=2$)
  computed from the initial condition of resting fluid,
  $\omega = 0$ and ellipsoid in verical position ($\theta = 0$), to
  $\theta \approx \nicefrac{\pi}{2}$. Local label stands for
  ELIBB
  and
  Tao methods, LIR means \emph{local iteration refilling}
  .\label{img:jeffery2d_convergence}}

\end{figure}

In figure~\ref{img:jeffery2d_convergence} we also computed the average torque squared fluctuation $\overline{TSF}$ on  $0<\theta\lesssim\nicefrac{\pi}{2}$ for a ellipsoidal cylinder and plotted it for an increasing resolution of the ellipsoids. The results show convergence with approximate slope $x^{-4}$ (this is expected for second-order schemes since we are working on the squared fluctuations).

To summarize the results of this section, the ELIBB shows good stability propriety in the simulation of a fluid-rigid body problem interaction. The ELIBB leads to results comparable with the BFL for the ellipsoid dynamics, both in terms of angular velocity evolution and in terms of torque fluctuation. Nevertheless, for the local methods tested the refill algorithm is important to reduce the effects on the torque due to the pressure oscillation.

\section{Conclusions}
In this paper, we presented a novel class of enhanced single-node boundary conditions (ELIBB). The \emph{enhancement} theoretically derives from the introduction of wall populations and the optimization of coefficients correlated with the quality of the interpolation in equation~\eqref{eq:ELIBB-general}. The resulting boundary methods are \emph{single-node}, which means that they are suitable to simulate complex shapes with narrow gaps containing only one lattice node, without introducing special conditions. This \emph{physical locality} feature of the ELIBB, facilitates the management of complex geometries. Thus, the ELIBB can be implemented in \emph{algorithmic local} way without any need of introducing special cases near corners or narrow gaps. In parallel simulations no data will need to be recovered from neighboring cells, improving in this way parallel computing performances. Therefore, ELIBB boundary conditions are attractive for GPU based implementations.

The novel boundary conditions show, in the investigated experiments, a robust behavior in terms of accuracy of the velocity field at low resolutions and large values of the relaxation parameters $\tau$, and, in this range, they appear to be more accurate than the well-established Bouzidi-Firdaouss-Lallemand~\citep{bouzidi2001momentum} boundary condition. The ELIBB variants shares with other interpolated bounce-back methods some limitations: mass violation and viscosity dependence. In the performed experiment, mass violation of the novel methods is higher compared to the BFL at low resolutions. At higher resolutions, this gap is bridged and the mass violation becomes similar to the one of BFL. Further investigations are needed to quantify these preliminary findings.

We showed that the method recently proposed by Tao \emph{et al.}~\citep{tao2018onepoint}, can be interpreted as a variant of the ELIBB family. Nonetheless, the variants proposed in the present paper show a non-negligible improvement with respect to the Tao \emph{et al.}~\citep{tao2018onepoint} method, at least for the low Reynolds number regime.

The novel class of boundary conditions is suitable to describe moving boundaries immersed in the fluid. In particular, it proved to be stable and well-behaving when describing the dynamics of a rigid body in a shear Besides, \emph{local iteration refill} algorithm has proved to be a good companion for the ELIBB. After its adoption to reduce pressure oscillations, single-node techniques and the BFL method showed the same level of noise on the torque acting on the rigid body.

We additionally provided an alternative non-equilibrium estimation scheme that can be valuable, in perspective when a more advanced boundary modeling is needed in place of a simple Dirichlet boundary condition (\emph{e.g.} wall models for turbulent flows). This aspect needs further investigation in the future.

From our first initial results, the ELIBB display encouraging improvements over some earlier single-nodes boundary conditions, yet maintaining an easy implementation characteristic. Therefore it is a good general-purpose candidate to replace the BFL in many applications where a single-node method is required. Nevertheless, more systematic studies are necessary to properly compare it with other techniques, especially in the range of higher Reynolds numbers.

\begin{acknowledgments}
  Fruitful discussions with Christophe Coreixas and Christos Kotsalos are gratefully acknowledged. This project has received funding from the European Union's Horizon 2020 research and innovation programme under grant agreement No 823712 (CompBioMed2 project).
\end{acknowledgments}

\appendix
\section{Hermite Polynomials}
\label{sec:hermite}
The Hermite functions are the solution solutions to the Hermite ordinary differential equation~\citep{arfken_chapter_2013}:
\begin{equation}
  \mathsf{H}_{n}^{\prime \prime}(x)-2 x \mathsf{H}_{n}^{\prime}(x)+2 n \mathsf{H}_{n}(x)=0\, .\label{eq:hermite_ode}
\end{equation}
The functions $\mathsf{H}_{n}$ are of integer degree $n$ solution of~\eqref{eq:hermite_ode}
are the \emph{physicist} Hermite polynomials~\citep{abramowitz_handbook_1965}.
Their general expression can be evaluated using the Rodriguez formula~\citep{arfken_chapter_2013}
\begin{equation}
  \mathsf{H}_n(x) = (-1)^n e^{x^2}\frac{d^n}{dx^n}e^{-x^2}.
\end{equation}
or, equivalently, using the generating function~\citep{arfken_chapter_2013,arfken_chapter_2013_12}
\begin{equation}
  g(x, t)=e^{-t^{2}+2 t x}=\sum_{n=0}^{\infty} \mathsf{H}_{n}(x) \frac{t^{n}}{n !}
\end{equation}
In the LB context is more common to refer to the \emph{probabilistic}
Hermite polynomials, which are a rescaled version of the $\mathsf{H}_{n}(x)$. In one dimension
their expression is~\citep{abramowitz_handbook_1965}
\begin{equation}
  H_{n}(x) =2^{-\frac{n}{2}} \mathsf{H}_{n}\left(\frac {x}{\sqrt 2} \right) = (-1)^n e^{\frac{x^2}{2}}\frac{d^n}{dx^n}e^{-\frac{x^2}{2}}.
\end{equation}
For the polynomial expansion of the probability density function in the velocity space
it is necessary to perform a multivariate extension of the classic Hermite polynomial~\citep{grad_kinetic_1949}.
In this case the Rodrigues' formulas reads
\begin{equation}
  \bm{H}_{n}(\bm{x})=H_{\alpha_{1}\ldots\alpha_{n}}(\bm{x})=(-1)^{n}e^{\frac{\bm{x}^{2}}{2}}\frac{\partial^{n}}{\partial_{\alpha_{1}\ldots\alpha_{n}}}e^{-\frac{\bm{x}^{2}}{2}}.
\end{equation}
where $\bm{x} = (x_1,\ldots,x_d)$, $d$ is the dimensionality and $\alpha_i\in\left\{x_1,\ldots,x_d\right\}$.
In the velocity space the expression of the first polynomials reads:
\begin{equation}
  \begin{aligned}
    H & =1\\
    H_{\alpha_{1}} & =\var_{\alpha_{1}}\\
    H_{\alpha_{1}\alpha_{2}} & =\var_{\alpha_{1}}\var_{\alpha_{2}}-\delta_{\alpha_{1}\alpha_{2}}\\
    H_{\alpha_{1}\alpha_{2}\alpha_{3}} & =\var_{\alpha_{1}}\var_{\alpha_{2}}\var_{\alpha_{3}}\\
    & -\left(\delta_{\alpha_{1}\alpha_{2}}\var_{\alpha_{3}}+\delta_{\alpha_{1}\alpha_{3}}\var_{\alpha_{2}}+\delta_{\alpha_{2}\alpha_{3}}\var_{\alpha_{1}}\right)\\
    H_{\alpha_{1}\alpha_{2}\alpha_{3}\alpha_{4}} & =\var_{\alpha_{1}}\var_{\alpha_{2}}\var_{\alpha_{3}}\var_{\alpha_{4}}\\
    & -\left(\delta_{\alpha_{1}\alpha_{2}}\var_{\alpha_{3}}\var_{\alpha_{4}}+\delta_{\alpha_{1}\alpha_{3}}\var_{\alpha_{2}}\var_{\alpha_{4}}+\delta_{\alpha_{1}\alpha_{4}}\var_{\alpha_{2}}\var_{\alpha_{3}}\right.\\
    & +\left.\delta_{\alpha_{2}\alpha_{3}}\var_{\alpha_{1}}\var_{\alpha_{4}}+\delta_{\alpha_{2}\alpha_{4}}\var_{\alpha_{1}}\var_{\alpha_{3}}+\delta_{\alpha_{3}\alpha_{4}}\var_{\alpha_{1}}\var_{\alpha_{2}}\right)\\
    & +\left(\delta_{\alpha_{1}\alpha_{2}}\delta_{\alpha_{3}\alpha_{4}}+\delta_{\alpha_{1}\alpha_{3}}\delta_{\alpha_{2}\alpha_{4}}+\delta_{\alpha_{1}\alpha_{4}}\delta_{\alpha_{2}\alpha_{3}}\right)
  \end{aligned}
\end{equation}

In the discrete case, the Hermite polynomial needs to be rescaled as a consequence of the rescaling of the discrete
velocities $\bm c_i = \bm \xi_i c_s$
\renewcommand{\var}{c}
\begin{align}
  \mathcal{H}_{i} & =1\\
  \mathcal{H}_{{\scriptscriptstyle i,\alpha_{1}}} & =\var_{i,\alpha_{1}}\\
  \mathcal{H}_{{\scriptscriptstyle i,\alpha_{1}\alpha_{2}}} & =\var_{i,\alpha_{1}}\var_{i,\alpha_{2}}-c_{s}^{2}\delta_{\alpha_{1}\alpha_{2}}\\
  \mathcal{H}_{{\scriptscriptstyle i,\alpha_{1}\alpha_{2}\alpha_{3}}} & =\var_{i,\alpha_{1}}\var_{i,\alpha_{2}}\var_{i,\alpha_{3}}\\
  & -c_{s}^{2}\left(\delta_{\alpha_{1}\alpha_{2}}\var_{i,\alpha_{3}}+\delta_{\alpha_{1}\alpha_{3}}\var_{i,\alpha_{2}}+\delta_{\alpha_{2}\alpha_{3}}\var_{i,\alpha_{1}}\right)\\
  \mathcal{H}_{{\scriptscriptstyle i,\alpha_{1}\alpha_{2}\alpha_{3}\alpha_{4}}} & =\var_{i,\alpha_{1}}\var_{i,\alpha_{2}}\var_{i,\alpha_{3}}\var_{i,\alpha_{4}}\\
  & -c_{s}^{2}\left(\delta_{\alpha_{1}\alpha_{2}}\var_{i,\alpha_{3}}\var_{i,\alpha_{4}}+\delta_{\alpha_{1}\alpha_{3}}\var_{i,\alpha_{2}}\var_{i,\alpha_{4}}\right.\\
  & +\delta_{\alpha_{1}\alpha_{4}}\var_{i,\alpha_{2}}\var_{i,\alpha_{3}}+\delta_{\alpha_{2}\alpha_{3}}\var_{i,\alpha_{1}}\var_{i,\alpha_{4}}\\
  & +\left.\delta_{\alpha_{2}\alpha_{4}}\var_{i,\alpha_{1}}\var_{i,\alpha_{3}}+\delta_{\alpha_{3}\alpha_{4}}\var_{i,\alpha_{1}}\var_{i,\alpha_{2}}\right)\\
  & +c_{s}^{4}\left(\delta_{\alpha_{1}\alpha_{2}}\delta_{\alpha_{3}\alpha_{4}}+\delta_{\alpha_{1}\alpha_{3}}\delta_{\alpha_{2}\alpha_{4}}+\delta_{\alpha_{1}\alpha_{4}}\delta_{\alpha_{2}\alpha_{3}}\right)
\end{align}
where $\mathcal{H}_{{\scriptscriptstyle i,\alpha_{1}..\alpha_{n}}}=\mathcal{H}_{{\scriptscriptstyle \alpha_{1}..\alpha_{n}}}(\bm c_i)$ and consequently $\bm{\mathcal{H}}_{i,n}=\bm{\mathcal{H}}_n\left( \bm c_i \right)$.

\section{Regularization procedures}
\label{sec:reg}
The concept of regularized populations was originally introduced by Skordos~\cite{SKORDOS_PRE_49_1993} and further developed by Latt, Malaspinas \emph{et al.}~\citep{latt2006lattice,latt2007hydrodynamic,malaspinas2009lattice,malaspinas2012consistent,Malaspinas2015}. The cornerstone of this approach is to reconstruct populations keeping only the minimal information to ensure that the correct macroscopic behavior is recovered. The purpose is to discard the information contained in the population that does not refer to the Navier-Stokes physics, to filter numerical errors. To this effect, populations are reconstructed from a truncated Hermite polynomial expansion of $f^{(0)}_i$ and $f^{(1)}_i$ at order $M$ and $N$ respectively~\cite{grad_kinetic_1949,SHAN_PRL_80_1998,shan_kinetic_2006}. It is possible to exemplify the regularized LB considering the following steps:
\begin{enumerate}
  \item Consider the initial condition, at iteration $t-1$, in which the non-equilibrium populations
  consists solely of hydrodynamic components $f_{i}^\neqq = f_{i}^{(1)}\sim \mathcal{O}(\mathrm{Kn}^1 f^{\eq})$ with null
  high order Hermite polynomials components. Where higher order means --not necessary to recover the Navier-Stokes level physics--. We have,
  \begin{subequations}
  \begin{equation}
            f_{i}=f_{i,M}^{{(0)}}+f_{i,N}^{(1)}\ ,
  \end{equation}
    \begin{equation}
      f_{i}^{(0)}=f_{i,M}^{(0)}\ ,
    \end{equation}
    \begin{equation}
      f_{i}^{(1)}=f_{i,N}^{(1)}\ ,
    \end{equation}
  \end{subequations}
  where $M$ and $N$ are the polynomial orders. And, after collision, the populations reads
  \begin{equation}
    f_{i}^{*}=f_{i,M}^{{(0)}}+\left(1-\frac{1}{\tau}\right)f_{i,N}^{(1)}
  \end{equation}
  where $f_{i}^{*}$ denotes a post-collision population.
  \item After streaming, at time $t$, non-hydrodynamic error components $\varepsilon_{i}$ caused by the numerical discretization appear
  \begin{equation}
    \tilde{f}_{i}=\overbrace{f_{i,M}^{(0)}\underbrace{\cancel{+\varepsilon_{i}^{\eq}}}_{\mathclap{\text{to be filtered}}}}^{\mathclap{\tilde{f}^{{\rm eq}}}}\ +\ \overbrace{f_{i,N}^{(1)}\underbrace{\cancel{+\varepsilon_{i}^{\neqq}}}_{\mathclap{\text{to be filtered}}}}^{\mathclap{\tilde{f}^{{\rm neq}}}}
  \end{equation}
  where $\tilde{f}_i$, $\tilde{f}_i^\eq$, and $\tilde{f}_i^\neqq$ are respectively the non-filtered population, equilibrium component and
  non-equilibrium component.
  \item Projective regularization step: it is a filtering process consisting in estimating $f_{i,M}^{(0)}$
  and  $f_{i,N}^{(1)}$ from the Hermite moments
  of $\tilde{f}^{{\rm eq}}_i$ and $\tilde{f}^{{\rm neq}}_i$ discarding high order moments contributions.
   The projective
  regularization step reads
  \begin{equation}\label{eq:f_expansion}
    f_{i,N}^{(1)}\approx
    f_{i,N}^{\neqq}=w_{i}\sum_{n=0}^{N}\frac{1}{n!c_{s}^{2n}}\tilde{\boldsymbol{a}}^{\neqq}_{n}:\bm{\mathcal{H}}_{i,n}
  \end{equation}
  \begin{equation}
    \tilde{\boldsymbol{a}}^{\neqq}_n\approx\sum_{i=0}^{Q-1} w_i \left(\tilde f_{i}-f_{i,M}^{{\rm eq}}\right)\bm{\mathcal{H}}_{i,n}
    \label{eq:RBGKan}
  \end{equation}
\begin{equation}
  f_{i,M}^{\rm eq} = f_{i,M}^{(0)} =w_{i}\sum_{n=0}^{M}\frac{1}{n!c_{s}^{2n}}{\boldsymbol{a}}^{(0)}_{n}:\bm{\mathcal{H}}_{i,n}
\end{equation}
  where $\bm{\mathcal{H}}_{i,n}=\bm{\mathcal{H}}_{n}(\bm{c}_{i}\!\!=\!\!c_s\bm{ \xi_i})$ is a Hermite polynomial of   order $n$ (see \cref{sec:hermite}), $\bm a_{n}$ is its corresponding tensor of coefficients, ``:'' is the Frobenius   inner product~\citep{amir-moez_generalized_1960}, $w_i$ are the quadrature weights and  $c_s$ is the lattice constant.

  \item finally, filtered population $f_i$ is reconstructed as
  \begin{equation}
    f_i \approx f^{(0)}_{i,M} + f_{i,N}^{\neqq}.
  \end{equation}
\end{enumerate}
Even if the RBGK was successful, it does not filters out all the non-hydrodynamic components, because the $f_i^{\neqq}$ computed from the projection still contains numerical errors due to higher perturbation order (in Knudsen number) components $\tilde{f}_i^{(2)}+\tilde{f}_i^{(3)}+\ldots\ $. For this reason, it has been extended by the recursive regularized model (RRBGK). In this case, the Hermite coefficients appearing in equations~\eqref{eq:RBGKan} are recomputed using the recursive formulas derived in references~\citep{malaspinas2009lattice,malaspinas2012consistent,Malaspinas2015} (see equations~\eqref{eq:shan} and \eqref{eq:malaspinas} in \cref{sec:hydro_limit}). The recursive formulation improves the filtering of numerical non-hydrodynamics errors for the Hermite components of order higher than two.

\section{Recursive estimation of the non-equilibrium populations}
\label{sec:hydro_limit}

In this section, we present the formal techniques used to enhance the boundary condition with an estimation of the wall populations. To achieve this, it is necessary to formally unfold the relation that elapses between the mesoscopic representation of the Navier-Stokes equations and the mesoscopic perspective of the BE. Although diverse mathematical approaches exist, the most common in the LB community is the Chapman-Enskog perturbative expansion (CE)~\citep{chapman_mathematical_1953}. This method has been extended with the Grad-Hermite (GH) expansion~\citep{grad_kinetic_1949} leading to an elegant and systematic approach recovering the hydrodynamic limit of both the BE and LBE~\citep{shan_kinetic_2006,latt2007hydrodynamic,malaspinas2009lattice,Malaspinas2015,malaspinas2012consistent}.

In the Chapman-Enskog-Hermite expansion~\citep{chapman_mathematical_1953,malaspinas2009lattice,Malaspinas2015, malaspinas2012consistent} the populations and the LBE are expanded in a perturbative formulation and decomposed in a set of Hermite-basis moments equations. The perturbative expansion of the velocity distribution function is given by equation~\eqref{eq:CE}.
The equilibrium $f^{(0)}$ and first non-equilibrium component $f^{(1)}$ components of equation~\eqref{eq:CE} can be projected in
the Hermite polynomial basis
\begin{align}
  f_{i,M}^{\rm eq} = f_{i,M}^{(0)} & =w_{i}\sum_{n=0}^{M}\frac{1}{n!c_{s}^{2n}}{\boldsymbol{a}}^{(0)}_{n}:\bm{\mathcal{H}}_{i,n}\label{eq:feqHemite}\\
  f_{i,N}^{(1)} & =w_{i}\sum_{n=0}^{N}\frac{1}{n!c_{s}^{2n}}{\boldsymbol{a}}^{(1)}_{n}:\bm{\mathcal{H}}_{i,n}\label{eq:fneqHemite}
\end{align}
where $\bm a^{(0)}_{n}$ and $\bm a^{(1)}_{n}$ are $n$-th order Hermite tensor coefficients of respectively the \emph{zero} and the \emph{first} order of the CE expansion (equation~\eqref{eq:CE})
\begin{align}
  \boldsymbol{a}_{n}^{(0)} & =\sum_{i=0}^{Q-1}w_{i}f_{i}^{(0)}\bm{\mathcal{H}}_{i,n}\\
  \boldsymbol{a}_{n}^{(1)} & =\sum_{i=0}^{Q-1}w_{i}f_{i}^{(1)}\bm{\mathcal{H}}_{i,n}\ .
\end{align}
Performing the CE of the LBE give rise an hierarchy of equations. Taking the moments of the first two equations (order \emph{zero} and \emph{one} of the CE) and comparing them with the Navier-Stokes equations, one can derive the Malaspinas' recursive formulas~\citep{malaspinas2009lattice,malaspinas2012consistent,Malaspinas2015}, that in the isothermal case read respectively
\begin{equation}
  \rho a_{\alpha_{1}\twoldots\alpha_{n}}^{(0)}=a_{\alpha_{n}}^{(0)}a_{\alpha_{1}\twoldots\alpha_{n-1}}^{(0)}\quad \forall n\geq 2,
  \label{eq:shan}
\end{equation}
and
\begin{equation}
  \rho a_{\alpha_{1}\twoldots\alpha_{n}}^{(1)}=a_{\alpha_{n}}^{(0)}a_{\alpha_{1}\twoldots\alpha_{n-1}}^{(1)}+\sum_{l=1}^{n-1}a_{\alpha_{1} \twoldots \alpha_{l-1} \alpha_{l+1} \twoldots \alpha_{n-1}}^{(0)}a_{\alpha_{l}\alpha_{n}}^{(1)}
  \label{eq:malaspinas}
\end{equation}
$\forall n \geq 3\nonumber$. The comparison between the Hermite moments of the Chapman-Enskog expanded LBE and the Navier-Stokes equation allows also to bind the relaxation time and the fluid viscosity through the following relation
\begin{equation}
  a_{\alpha_1\alpha_2}^{(1)}=\tau\rho\Lambda_{\alpha_1\alpha_2}
  \label{eq:a12Eqlambda}
\end{equation}
where $\Lambda_{\alpha_{1}\alpha_{2}}=\partial_{\alpha_{1}}u_{\alpha_{2}}+\partial_{\alpha_{2}}u_{\alpha_{1}}$ is the macroscopic isothermal-incompressible stress tensor and $D$ is the problem dimensionality (one, two or three). 

In the RRBGK the recursive formulas are used in equations \eqref{eq:feqHemite} and \eqref{eq:fneqHemite} to rebuild the populations up to the first oreder of the CE expansion, filtering non-hydrodynamic components. To correctly recover a Hermite moment of order $n$ at the Knunsen order $k$, $M$ should be greater or equal to $n+k+1$ and $N\geq n+k$~\citep{shan_kinetic_2006}. Thus, even in the isothermal case, the optimal order of truncation for the equilibrium is the fourth.
Note that, even if in the case of isothermal lattices only moments up to the second polynomial order are correctly recovered, equation \eqref{eq:shan} allows to recompute $\bm{a}_{>2}^{(0)}$ from $\bm{a}_2^{(0)}$. Regarding the order of truncation $N$, the minimal value in the isothermal case is $N=3$, but there is evidence that the optimal value is $N=M=4$~\citep{COREIXAS_PhD_2018}. Here we show the procedure for the order three truncation for sake of conciseness. Using the Malaspinas' formula~\eqref{eq:malaspinas} in \eqref{eq:fneqHemite} truncated at the third order and recalling that equations~\eqref{eq:a12Eqlambda} one gets
\begin{multline}
  f_{i,3}^{(1)}\approx-w_{i}\frac{\tau\rho}{c_{s}^{2}}\left[\frac{1}{2c_{s}^{2}}{\cal H}_{i,\alpha_{1}\alpha_{2}}\Lambda_{\alpha_{1}\alpha_{2}}+\frac{1}{6c_{s}^{4}}{\cal H}_{i,\alpha_{1}\alpha_{2}\alpha_{3}}\right.\\
  \left.\left(\Lambda_{\alpha_{1}\alpha_{2}}u_{\alpha_{3}}+\Lambda_{\alpha_{1}\alpha_{3}}u_{\alpha_{2}}+\Lambda_{\alpha_{2}\alpha_{3}}u_{\alpha_{1}}\right)\vphantom{\frac{1}{6c_{s}^{4}}}\right],
  \label{eq:fi_neq}
\end{multline}
where we used the Einstein notation, therefore summation is implicit for repeated indexes, $\alpha_i\in \left\{x_1,\ldots,x_d \right\}$, $d$ is the dimensionality of the velocity space and $x_1,\ldots,x_d$ are the names of the space axes.

Thanks to the symmetry proprieties of the Hermite polynomials, from the previous equation follows the symmetry of non-equilibrium populations
\begin{equation}
  f_{i}^{\neqq}(x_{\alpha},t_{\alpha})=f_{\bar{i}}^{\neqq}(x_{\alpha},t_{\alpha}).\label{eq:neq_bb}
\end{equation}
The latter formula is the starting point of the wet-node Zou-He~\citep{he1995analysis,zou1997onpressure} boundary condition, also know as non-equilibrium bounce-back.
Some authors~\citep{tao2018onepoint,liu2019lattice} use this idea to compute off-lattice unknown non-equilibrium components in an approximate way
\begin{equation}
  f_{i}^{\neqq}(x_{\alpha},t_{\alpha})\approx f_{\bar{i}}^{\neqq}(x_{\beta}\approx x_{\alpha}+\delta_{x}x_{\alpha},t_{\beta}\approx t_{\alpha}+\delta_{t}t_{\alpha})\label{eq:approx_neq_bb}
\end{equation}
where $x_{\alpha}$ is the coordinate of the off-lattice position of interest, $x_\beta$ is the coordinate of a lattice node, $\delta_{x}$ is a small space distance, $\delta_{t}$
denotes a small time interval and $t_\beta$ is a timestep that can differ from  $t_\alpha$.
The approximated non-equilibrium bounce-back is a first-order approximation for the non-equilibrium component, nevertheless as shown by Chun \emph{et al.}~\citep{chun2007interpolated} the non-equilibrium component is a second-order correction over the equilibrium component. Thanks to this fact, equation (\ref{eq:approx_neq_bb}) can still be used to develop a second-order accurate boundary condition.

To be more general, in this paper we propose an additional approach to explicitly estimate the \emph{off-lattice} values of $f_i^{(1)}$. Namely, using the expression~\eqref{eq:fi_neq} to approximate the non-equilibrium component at the wall location. In equations~\eqref{eq:fi_neq} all the variables appearing should be interpreted in our case as computed at wall location. For the stresses we use a first-order approximation $\Lambda_{\alpha_1\alpha_2}(\bm x_W)\approx\Lambda_{\alpha_1\alpha_2}(\bm x_F)$. Adopting~\eqref{eq:fi_neq} can require a slightly higher computational cost, but it allows for more flexibility on the boundary modeling if needed. For example, when is necessary to model $\Lambda_{\alpha_1\alpha_2}(\bm x_W)$ as in references~\citep{malaspinas_wall_2014,park_improved_2014}.

\section{Technical details of Jeffery's orbit test-case}

\subsection{Qualitative approach to evaluate torque oscillations}
\label{sec:qualitative_approach}
The theoretical value of the torque acting on the ellipsoid (or ellipsoidal cylinder) changes over time. If the ellipsoid inertia is small, the magnitude of the torque oscillation exceeds the value of the theoretical torque. Unfortunately, the analytical value of the torque is not available and this makes it impossible to compute the oscillation because the baseline solution is unknown. To give a qualitative representation of the evolution of the torque squared fluctuation, we decided to compute a smooth numerical baseline solution using interpolating polynomials. This baseline regular numerical solution is then used to compute the squared fluctuation of the torque (TSF) over time. The TSF of the torque computed in this way is noisy and not enough precise to obtain quantitative results. Nevertheless, we can use it to perform a second interpolation with a polynomial of the same order to get a qualitative estimation of the TSF evolution over time and use it to compare visually different methods.

\begin{figure}[t]
  \centering
  \includegraphics[width=1.\columnwidth,valign=c]{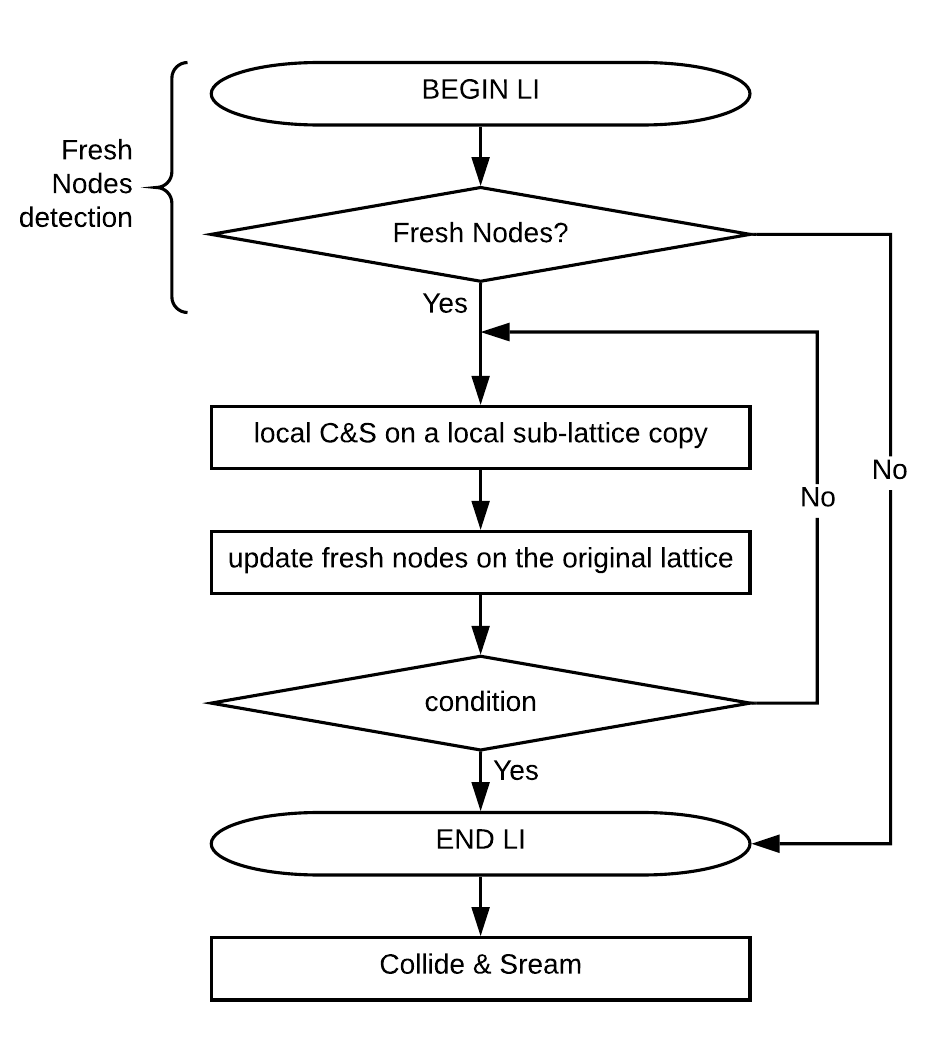}
  \caption{The modified LIR algorithm.}
  \label{fig:lir}
\end{figure}

In detail, the first polynomial least squares interpolation is of order $n=30$ and reads
\begin{equation}
  T_{\text{ip,n}}\left(t_{j}\right)={\rm argmin}\left(\sum_{j=0}^{k}\left|p_{n}\left(t_{j}\right)-T_{j}\left(t_{j}\right)\right|^{2}\right).
\end{equation}
After the computation of the torque squared fluctuation (TSF)
\begin{equation}
  TSF=\left(T-T_{\text{ip},n}\right)^{2},
\end{equation}
a second interpolating polynomial of the same order is computed for the
fluctuations
\begin{equation}
  TSF_{\text{ip,n}}\left(t_{j}\right)={\rm argmin}\left
  (\sum_{j=0}^{k}\left|p_{n}\left(t_{j}\right)-T_{{\rm ip},n,j}\left
  (t_{j}\right)\right|^{2}\right)
  \label{eq:TSF30}
\end{equation}
where $t_{j}$ is the $j-th$ iteration. This procedure allows to produce
regular and qualitative graphs of the torque squared fluctuations in time.

\subsection[]{{Local Iteration Refilling}}
\label{sec:LIR}
The Local Iteration refilling (LIR) proposed in~\citep{chen2014acomparative} is modified to make it consistent with a thin shell two-dimensional representation of the boundary. To do so, the LIR is applied \emph{before} the global collide and stream, and not \emph{after}. This detail guarantees that the streaming step does not move wrong populations outside the \emph{fresh nodes} before the application of the LIR. The modified algorithm is represented in figure~\ref{fig:lir}.

\bibliographystyle{myapsrev4-2}
\bibliography{boaundaryConditions}
\end{document}